\documentclass[journal,10 pt,twocolumn]{IEEEtranTCOM}
\usepackage{epsfig}
\usepackage{graphicx}
\usepackage{caption}
\usepackage{subcaption}
\usepackage{balance}
\usepackage{float}
\usepackage{psfrag}
\usepackage{amsfonts,amsmath,amssymb,mathrsfs,txfonts}

\newtheorem{lemma}{Lemma}

\newtheorem{definition}{Definition}

\newtheorem{remark}{Remark}
\usepackage{url}
\usepackage{cite}
\usepackage{verbatim}
\usepackage{multirow}
\usepackage{multicol}
\usepackage[font=small]{caption}
\usepackage{pifont}
\usepackage{algorithm,algpseudocode}
\usepackage{color}
\usepackage{lettrine}
\begin{document}
\title{Generalized Firefly Algorithm for Optimal Transmit Beamforming}
\author{Tuan Anh Le and Xin-She Yang
\thanks{T. A. Le and X.-S. Yang are with the Faculty of Science and Technology, Middlesex University, London, NW4 4BT, UK. Email: \{t.le; x.yang\}@mdx.ac.uk. }
\thanks{This paper has been presented in part at the IEEE Vehicular Technology Conference (VTC 2023-Spring), Florence, Italy, June, 20-23, 2023.}
}
\markboth{IEEE Transactions on Wireless Communications, DOI: 10.1109/TWC.2023.3328713 }%
{Shell \MakeLowercase{\textit{et al.}}: Bare Demo of IEEEtran.cls for IEEE Journals}
\maketitle

\begin{abstract}
This paper proposes a generalized Firefly Algorithm (FA) to solve an optimization framework having objective function and constraints as multivariate functions of independent optimization variables. Four representative examples of how the proposed generalized FA can be adopted to solve downlink beamforming problems are shown for a classic transmit beamforming, cognitive beamforming, reconfigurable-intelligent-surfaces-aided (RIS-aided) transmit beamforming, and RIS-aided wireless power transfer (WPT). Complexity analyzes indicate that in large-antenna regimes the proposed FA approaches require less computational complexity than their corresponding interior point methods (IPMs) do, yet demand a higher complexity than the iterative and the successive convex approximation (SCA) approaches do. Simulation results reveal that the proposed FA attains the same global optimal solution as that of the IPM for an optimization problem in cognitive beamforming. On the other hand, the proposed FA approaches outperform the iterative, IPM and SCA in terms of obtaining better solution for optimization problems, respectively, for a classic transmit beamforming, RIS-aided transmit beamforming and RIS-aided WPT.
\end{abstract}
\IEEEpeerreviewmaketitle
\section{Introduction}
\lettrine[findent=1pt]{{\textbf{T}}}{ransmit} beamforming problems are normally cast as optimization problems where beamforming vectors are optimization variables. Two fundamental optimization problems in transmit beamforming include: i) minimizing the total transmit power subject to signal-to-interference-plus-noise-ratio (SINR) constraints \cite{Farrokh,Ami,Yongwei,Wei11}; ii) maximizing the weakest SINR subject to a total power constraint \cite{Weidong,Schubert2004}. In fact, these two problems are equivalent \cite{Hammarwall,Bjornson}. A generalized version of the second problem is introduced in \cite{Bjornson} where the objective is to maximize an arbitrary utility function of SINRs, which is strictly increasing in every receiver's SINR, subject to a power constraint. The other variation of the second optimization problem is the sum rate maximization \cite{GanZheng08,Bhagavatula}. Furthermore, additional constraints can be introduced to these fundamental problems to capture other wireless communication applications. For instance, a soft-shaping interference constraint was added for cognitive radio scenarios \cite{Huang2010new,Yongwei2012} while a power transfer constraint was included for simultaneous-wireless-information-and-power-transfer  scenarios \cite{Clerckx19}. In addition, various metrics have been utilized to formulate downlink beamforming optimization problems such as secrecy capacity \cite{DerrickPT}, energy efficiency \cite{Derrick12}, data transmission reliability, data transmission security, and power transfer reliability \cite{le2017robust}.

Since the SINR is a non-convex quadratic function of the beamforming vectors, the two fundamental beamforming optimization problems are NP-hard and cannot be solved in polynomial time. Fortunately, exploiting the hidden convexity property of the SINR metric, an elegant framework was proposed in  \cite{Ami} to convert these two optimization problems into convex conic programming forms, which can be effectively solved by a standard interior point method (IPM). Furthermore, uplink-downlink duality was utilized to derive iterative algorithms to find optimal beamforming vectors for some power minimization problems, e.g., \cite{Farrokh,WeiYu07,Wei11,TuanTcom2014}. An iterative algorithm  was introduced in \cite{GanZheng08} to attain optimal beamforming vectors for the sum rate maximization.

Numerous transmit beamforming problems can be realized in quadratically constrained quadratic programs (QCQPs) of beamforming vectors, which are mostly non-convex \cite{Zhi,Huang2010new}. To solve a QCQP problem, a semidefinite relaxation technique \cite{Mats} is adopted in which
the original QCQP is converted to a convex semidefinite programming (SDP) with new optimization variables as beamforming matrices. If solving the transformed SDP yields a rank-one optimal beamforming matrix, then this optimal matrix is also the optimal solution to the original QCQP. Otherwise, an approximated solution to the original QCQP can be obtained by exploiting some rank-one approximations or the Gaussian randomize procedure \cite{Zhi}. Unfortunately, obtaining such solution requires further computational resources yet results in a sub-optimal solution.

Optimization variables for downlink beamforming problems may include different types of  beamforming vectors. For example, in a reconfigurable-intelligent-surface-aided (RIS-aided) communication system, see e.g., \cite{Peng2022,Gong2023} and references therein, the optimization variables are active beamforming vectors for the base station (BS) and a passive beamformimg vector for the RIS. The objective function and/or constraints for a RIS-aided communication system are functions of both active and passive beamforming vectors. These beamforming vectors are independent variables yet need to be jointly optimized making their problems non-convex. Widely adopted approaches for tackling such problems are to iteratively solve two sub-optimization problems, a.k.a., alternative optimization (AO) approach \cite{Peng2022}, or to approximate a non-convex using first-order Taylor expansion, a.k.a., successive convex approximation (SCA) \cite{Wu2020}. In an AO approach, each of these two sub-optimization problems, one variable is treated as a constant while solving for the other. These sub-optimization problems themselves are mostly in QCQP forms. Due to the inherent non-convexity character of the original and sub-optimization problems, the resulting active and passive beamforming vectors may not be the global solutions.  Whereas in a SCA approach, a lower (or upper) bounded solution is normally attained. 

IPMs, a.k.a., barrier methods, are gradient based algorithms being good at exploitation,\footnote{Exploitation is the ability of using any information from the problem of interest to form new solutions which are better than the current ones \cite{YangFA2008}.} a.k.a., intensification, hence, they are regarded as effective methods to solve convex optimization problems \cite{Boyd_convex}. Unfortunately, most of transmit beamforming problems are non-convex. Solving non-convex optimization problems requires algorithms having better exploration\footnote{Exploration is the ability of efficient exploring the search space to form new solutions with sufficient diversity and far from the existing ones \cite{YangFA2008}.} ability than that of the IPMs to avoid getting trapped in a local mode. Firefly algorithm (FA), i.e., a nature-inspired algorithm, possesses both exploitation and exploration abilities. Consequently, FA is a good candidate for solving non-convex downlink beamforming problems. FA is an easy-to-implement, simple, and flexible algorithm based on the flashing characters and behaviour of tropical fireflies \cite{YangFA2008}. FA was first developed and published by Xin-She Yang, respectively, in late 2007 and in 2008 \cite{YangFA2008,YangFA2009} for  optimization problems with objective and constrains being functions of a single optimization variable. Although FA has been widely applied to many applications \cite{YangFA2020}, there has not been any significant work investigating the application of FA in solving transmit beamforming problems. There were only two attempts to adopt FA for a throughput maximization problem in \cite{Yamanaka} and for a power minimization problem in \cite{Tuanvtc23}. As these two attempts only capture two fundamental transmit beamforming problems, it is not clear how FA can be adopted to solve other types of transmit beamforming problems. 
 
This paper takes a further step on implementing FA to solve a wider range of transmit beamforming optimization problems. The contributions of the paper can be summarized as follows. 
\begin{itemize}
    \item The paper proposes a generalized FA to find the optimal solution of an optimization framework where its objective function and constraints are multivariate functions of multiple independent optimization variables. The problems in \cite{Yamanaka} and \cite{Tuanvtc23} are only two special cases of the proposed generalized FA while the proposed generalized FA is capable of handling a larger range of transmit beamforming problems.
    \item The paper shows four representative examples of how the generalized FA can be adopted for solving transmit beamforming problems, i.e., a classic transmit beamforming approach, a cognitive beamforming approach, a RIS-aided beamforming approach, and RIS-aided wireless power transfer (WPT) approach. The applications of the proposed generalized FA are beyond these four examples which are only given to showcase how different types of beamforming problems can be handled by the generalized FA.
    \item For the sake of completeness and comparison, the iterative closed form or SDP forms of the under investigated beamforming approaches are represented. The paper analyzes and compares the complexities of the iterative or SDP and FA implementations of each beamforming approach.
    \item Simulations are carried out to evaluate the performances of the proposed FAs for the classic transmit beamforming, cognitive beamforming, RIS-aided, and RIS-aided WPT beamforming approaches.
\end{itemize}

\emph{\textbf{Notation}:}
Lower and upper case letter $y$ and $Y$: a scalar; bold lower case letter $\mathbf{y}$: a column vector; bold upper case letter $\mathbf{Y}$: a matrix; $\left\|\cdot\right\|$: the Euclidean norm; $(\cdot)^T$: the transpose operator; $(\cdot)^H$: the complex conjugate transpose operator; $\textrm{Tr}\left(\cdot\right)$: the trace operator; $\mathbf{Y}\succeq \mathbf{0}$: $\mathbf{Y}$ is positive semidefinite; $\mathbf{I}_x$: an $x \times x$ identity matrix;   $\mathcal{O}$: the big O notation;  $\mathbb{C}^{M\times 1}$: the set of all $M\times 1$ vectors with complex elements; $\mathbb{H}^{M\times M}$: the set of all $M\times M$ Hermitian matrices; $y\sim\mathcal{CN}(0,\sigma^2)$: $y$ is a zero-mean circularly symmetric complex Gaussian random variable with variance $\sigma^2$;    $\textrm{diag}\left( \mathbf{y}\right)$: a diagonal matrix whose diagonal elements are the entries of vector $\mathbf{y}$; and finally $\textrm{diag}\left( \mathbf{Y}\right)$: a vector whose entries are the diagonal elements of matrix $\mathbf{Y}$.
\section{Generalized Firefly Algorithm Framework}
\subsection{Proposed Generalized Firefly Algorithm Framework}
The FA was developed based on the following three idealized rules \cite{YangFA2008,YangFA2009}. First, any firefly attracts other fireflies regardless of its sex. Second, the attractiveness of any firefly to the other one is proportional to its brightness. Both attractiveness and brightness decrease as the distance between these two fireflies increases. Given two flashing fireflies, the darker firefly will move towards the brighter one. If a firefly does not find any brighter one, it will make a random move. Third, the brightness of a firefly depends on the landscape of the objective function.

In this section, we propose a generalized FA to find optimal solution for an optimization framework containing both objective and constraints as multivariate functions of independent variables. To that end, we first introduce the following optimization framework.
\begin{equation}
\begin{aligned}\label{framework}
& \underset{\mathbf{A},\mathbf{B},\cdots,\mathbf{Z}}{\textrm{minimize}} & & f\left( \mathbf{A},\mathbf{B},\cdots,\mathbf{Z}\right), \\
& \mbox{subject to}\ & &  g_l\left( \mathbf{A},\mathbf{B},\cdots,\mathbf{Z}\right) \leq 0,\ l \in \{1,2,\dots, L\},\\
&&& h_k\left( \mathbf{A},\mathbf{B},\cdots,\mathbf{Z}\right) =0, \ k \in \{1,2,\dots K\},
 \end{aligned}
\end{equation}
where $\mathbf{A}\in \mathbb{C}^{M_a\times N_a}, \mathbf{B}\in \mathbb{C}^{M_b\times N_b},\cdots,\mathbf{Z}\in \mathbb{C}^{M_z\times N_z}$, i.e., $M_a, N_a, M_b,N_b,\cdots,M_z,N_z \geq 1$, are decision variables, a.k.a., optimization variables. Depending on the the values of $\left\{M_a, N_a, M_b,N_b,\cdots,M_z,N_z\right\}$, the decision variables can be matrices, vectors, scalars, or the combination of all. 

We continue by using the penalty method \cite{YangFA2008,YangFA2009} to equivalently rewrite \eqref{framework} as:
\begin{equation}
\begin{aligned}\label{framework_01}
& \underset{\mathbf{A},\mathbf{B},\cdots,\mathbf{Z}}{\textrm{minimize}} & & f\left( \mathbf{A},\mathbf{B},\cdots,\mathbf{Z}\right)+P\left( \mathbf{A},\mathbf{B},\cdots,\mathbf{Z}\right),
 \end{aligned}
\end{equation}
where $P\left( \mathbf{A},\mathbf{B},\cdots,\mathbf{Z}\right)$ is the penalty term defined as:
\begin{eqnarray}
P\left( \mathbf{A},\mathbf{B},\cdots,\mathbf{Z}\right)=\sum_{l=1}^L \lambda_l  \textrm{max}\{0,  g_l\left( \mathbf{A},\mathbf{B},\cdots,\mathbf{Z}\right)\}^2\nonumber\\+\sum_{k=1}^K \rho_k  \{  h_k\left( \mathbf{A},\mathbf{B},\cdots,\mathbf{Z}\right)\}^2.\label{penalty_term}
\end{eqnarray}
In \eqref{penalty_term}, $\lambda_l >0, \  \forall l,$ and $\rho_k >0, \  \forall k$, are penalty constants. Let $\{ \mathbf{A}_i,\mathbf{B}_i,\cdots,\mathbf{Z}_i\}$ be the $i$-th firefly amongst the population of $N$ fireflies, i.e., $i \in \{1,2,\cdots, N\}$. Following the second rule of the FA, the brightest firefly is the most attractive one. Since the proposed optimization framework is a minimization, we define the brightness of firefly $i$ as:\footnote{Note that if \eqref{framework} is a maximization problem, then \eqref{framework_01} can be expressed as: $\underset{\mathbf{A},\mathbf{B},\cdots,\mathbf{Z}}{\textrm{minimize}}~-~f\left( \mathbf{A},\mathbf{B},\cdots,\mathbf{Z}\right)+P\left( \mathbf{A}_i,\mathbf{B}_i,\cdots,\mathbf{Z}_i\right)$.}
\begin{equation}
    I_i\left( \mathbf{A}_i,\mathbf{B}_i,\cdots,\mathbf{Z}_i\right)=\frac{1}{f\left( \mathbf{A}_i,\mathbf{B}_i,\cdots,\mathbf{Z}_i\right)+P\left( \mathbf{A}_i,\mathbf{B}_i,\cdots,\mathbf{Z}_i\right)}.\label{light0}
\end{equation}
For any two fireflies $i,j\in\{1,2,\cdots,N \}$, if $ I_j\left( \mathbf{A}_j,\mathbf{B}_j,\cdots,\mathbf{Z}_j\right)~>~I_i\left( \mathbf{A}_i,\mathbf{B}_i,\cdots,\mathbf{Z}_i\right)$, then firefly $i$ will move towards firefly $j$ at $(n+1)$-th generation as:
\begin{eqnarray}
    \mathbf{A}_i^{(n+1)}&=&\mathbf{A}_i^{(n)}+\beta_{a,0}e^{-\gamma_x (r_{a,ij}^{(n)})^2}\left( \mathbf{A}_j^{(n)}-\mathbf{A}_i^{(n)}\right)+\alpha_a^{(n)} \pmb{\Lambda}_{a,i}^{(n)},\label{x}\\
    \mathbf{B}_i^{(n+1)}&=&\mathbf{B}_i^{(n)}+\beta_{b,0}e^{-\gamma_y (r_{b,ij}^{(n)})^2}\left( \mathbf{B}_j^{(n)}-\mathbf{B}_i^{(n)}\right)+\alpha_b^{(n)} \pmb{\Lambda}_{b,i}^{(n)},\label{y}\\
    \vdots \nonumber \\
    \mathbf{Z}_i^{(n+1)}&=&\mathbf{Z}_i^{(n)}+\beta_{z,0}e^{-\gamma_z (r_{z,ij}^{(n)})^2}\left( \mathbf{Z}_j^{(n)}-\mathbf{Z}_i^{(n)}\right)+\alpha_z^{(n)} \pmb{\Lambda}_{z,i}^{(n)},\label{z}
\end{eqnarray}
where $r_{a,ij}^{(n)}=|| \mathbf{A}_j^{(n)}-\mathbf{A}_i^{(n)}||, r_{b,ij}^{(n)}=|| \mathbf{B}_j^{(n)}-\mathbf{B}_i^{(n)}||,\cdots,r_{z,ij}^{(n)}=|| \mathbf{Z}_j^{(n)}-\mathbf{Z}_i^{(n)}||$  are the Cartesian distances which are not necessary Euclidean distances yet they can be any measure effectively characterized the quantities of interest in the optimization problem;  $\beta_{a,0},\beta_{b,0},\cdots,\beta_{z,0}$ are, respectively, the attractiveness at $r_{a,ij}^{(n)}=0, r_{b,ij}^{(n)}=0,\cdots,r_{z,ij}^{(n)}=0$; finally $\gamma_a,\gamma_b,\cdots,\gamma_z$ present the variations of the attractiveness. The second terms in \eqref{x},  \eqref{y}, and \eqref{z} capture the attractions. The third terms in \eqref{x},  \eqref{y}, and \eqref{z} are randomizations with randomization factors $\alpha_a^{(n)},\alpha_b^{(n)},\cdots,\alpha_z^{(n)}$ and $\pmb{\Lambda}_{a,i}^{(n)}\in\mathbb{C}^{M_a \times N_a},\pmb{\Lambda}_{b,i}^{(n)}\in\mathbb{C}^{M_b \times N_b},\cdots,\pmb{\Lambda}_{z,i}^{(n)}\in\mathbb{C}^{M_z \times N_z}$ being matrices of random numbers drawn from a Gaussian or an uniform distribution. The proposed generalized FA for solving the optimization framework \eqref{framework} is summarized in Algorithm~\ref{FAoriginal}, where $T$ is the maximum generation of the algorithm. For any particular optimization problem subsumed under the framework, the corresponding FA will have the same steps as those in Algorithm~\ref{FAoriginal} except the input, step 3, step 16, step 18, step 19, and the return value. 
\begin{algorithm}
	\caption{Generalized Firefly Algorithm for solving \eqref{framework}}
	\label{FAoriginal}
	\begin{algorithmic}[1]
		\State \textbf{Input:} { \it FA parameters}: $N$, $T$, $\lambda_t$, $\rho_k$, $\beta_{a,0},\beta_{b,0},\cdots,\beta_{z,0}$, $\gamma_a,\gamma_b,\cdots,\gamma_z$; {\it Optimization data:} the structures/parameters of functions $f\left( \mathbf{A},\mathbf{B},\cdots,\mathbf{Z}\right)$, $g_l\left( \mathbf{A},\mathbf{B},\cdots,\mathbf{Z}\right)$, $h_k\left( \mathbf{A},\mathbf{B},\cdots,\mathbf{Z}\right)$;
		\State Randomly generate $N$ populations $\{ \{ \mathbf{A}_1,\mathbf{B}_1,\cdots,\mathbf{Z}_1\}, \{ \mathbf{A}_2,\mathbf{B}_2,\cdots,\mathbf{Z}_2\},\cdots, \{ \mathbf{A}_N,\mathbf{B}_N,\cdots,\mathbf{Z}_N\}\}$;
		\State Evaluate the light intensities of $N$ population as \eqref{light0};
		\State  Rank the fireflies in a descending order of $I_i\left(\mathbf{A}_i,\mathbf{B}_i,\cdots,\mathbf{Z}_i\right)$;
		\State Define the current best solution: $I^{\star}:=I_1\left(\mathbf{A}^{\star},\mathbf{B}^{\star},\cdots,\mathbf{Z}^{\star}\right)$; $\{ \mathbf{A}^{\star},\mathbf{B}^{\star},\cdots,\mathbf{Z}^{\star}\}:=\{ \mathbf{A}_1,\mathbf{B}_1,\cdots,\mathbf{Z}_1\}$;
		\For{$n =1:T$}
		    \For{$i=1:N$}
		    \For{$j=1:N$}
		    \If{$I_i\left(\mathbf{A}_i,\mathbf{B}_i,\cdots,\mathbf{Z}_i\right)>I^{\star}$} 
		    \State $I^{\star}:=I_i\left(\mathbf{A}_i,\mathbf{B}_i,\cdots,\mathbf{Z}_i\right)$; $\{ \mathbf{A}^{\star},\mathbf{B}^{\star},\cdots,\mathbf{Z}^{\star}\}:=\{ \mathbf{A}_i,\mathbf{B}_i,\cdots,\mathbf{Z}_i\}$;
		    \EndIf
		    \If{$I_j\left(\{ \mathbf{A}_j,\mathbf{B}_j,\cdots,\mathbf{Z}_j\}\right)>I^{\star}$} 
		    \State $I^{\star}:=I_j\left(\mathbf{A}_j,\mathbf{B}_j,\cdots,\mathbf{Z}_j\right)$; $\{ \mathbf{A}^{\star},\mathbf{B}^{\star},\cdots,\mathbf{Z}^{\star}\}:=\{ \mathbf{A}_j,\mathbf{B}_j,\cdots,\mathbf{Z}_j\}$;
		    \EndIf
		    \If{$I_j\left(\mathbf{A}_j,\mathbf{B}_j,\cdots,\mathbf{Z}_j\right)>I_i\left(\mathbf{A}_i,\mathbf{B}_i,\cdots,\mathbf{Z}_i\right)$}
		\State Move firefly $i$ towards firefly $j$ as \eqref{x}-\eqref{z};
		    \EndIf
		\State Attractiveness varies with distances via $e^{-\gamma_a \left(r_{a,ij}^{(n)}\right)^2}, e^{-\gamma_b \left(r_{b,ij}^{(n)}\right)^2},\cdots,e^{-\gamma_z \left(r_{z,ij}^{(n)}\right)^2}$;
		\State Evaluate new solutions and update light intensity as \eqref{light0};
		    \EndFor 
		    \EndFor 
		    \State Rank the fireflies in a descending order  of $I_i\left(\mathbf{A}_i,\mathbf{B}_i,\cdots,\mathbf{Z}_i\right)$;
		    \State Update the current best solution: $I^{\star}:=I_1\left( \mathbf{A}^{\star},\mathbf{B}^{\star},\cdots,\mathbf{Z}^{\star}\right)$; $\{ \mathbf{A}^{\star},\mathbf{B}^{\star},\cdots,\mathbf{Z}^{\star}\}:=\{ \mathbf{A}_1,\mathbf{B}_1,\cdots,\mathbf{Z}_1\}$;
		\EndFor
		\State \Return $\{ \mathbf{A}^{\star},\mathbf{B}^{\star},\cdots,\mathbf{Z}^{\star}\}$.
		
	\end{algorithmic}
\end{algorithm}
\subsection{Asymptotic Convergence and Optimality}
Since the firefly algorithm, like quite a few other nature-inspired algorithms, is a metaheuristic algorithm, there is no rigorous proof of convergence so far in the current literature, despite many applications of such metaheuristic algorithms. In this section, we provide some intuitive discussions on the optimality and convergence of the FA framework.\footnote{Mathematical analysis of the FA's optimality and convergence deserves an important research topic. Such analysis is postponed to future research due to the space constraint. }

\subsubsection{Asymptotic Optimality} 
Without loss of generality, let $\gamma_a=\gamma_b=\cdots=\gamma_z=\gamma$, we consider two special cases of the variations of the attractiveness when $\gamma \rightarrow 0$ and $\gamma \rightarrow \infty$. When $\gamma \rightarrow 0$, it is clear that $e^{-\gamma (r_{a,ij}^{(n)})^2} \rightarrow 1, e^{-\gamma (r_{b,ij}^{(n)})^2} \rightarrow 1, \cdots, e^{-\gamma (r_{z,ij}^{(n)})^2} \rightarrow 1$. Therefore the attractivenesses in \eqref{x},  \eqref{y}, and \eqref{z} are constant and, respectively, equal to $\beta_{a,0}, \beta_{b,0}$, and $\beta_{z,0}$. Equivalently, it is an idealized sky scenario where the brightness of each firefly does not change over the distance, which can be seen everywhere. Consequently, a global optimum can be obtained.

On the other hand, when  $\gamma \rightarrow \infty$, it is obvious that $e^{-\gamma (r_{a,ij}^{(n)})^2} \rightarrow 0, e^{-\gamma (r_{b,ij}^{(n)})^2} \rightarrow 0, \cdots, e^{-\gamma (r_{z,ij}^{(n)})^2} \rightarrow 0$, indicating that the attractiveness of each firefly is zero. Equivalently, each firefly is randomly in a heavily foggy region and cannot be seen by the others. Each will randomly move and the optimality is not always guaranteed. In this case, FA is equivalent to a random search approach.

In fact, the attractiveness is in between these two extreme cases, i.e., $0 < \gamma <\infty$. The value of $\gamma^{-0.5}$ defines the average distance of a herd of fireflies being seen by its adjacent herds. Hence, the entire population can be separated into number of herds. This automatic division property provides FA suitable ability of handling highly nonlinear and multimodal optimization problems. By controlling the attractiveness $\gamma_a, \gamma_b, \cdots, \gamma_z$ and the roaming randomness $\alpha_a, \alpha_b, \cdots, \alpha_z$, it has been shown in previous studies that FA can outperform both Particle Swarm Optimization (PSO), see, e.g., \cite{Iztok,Yang2018,Windarto,w15101906}, and random search approaches, see e.g., \cite{YangFA2008,YangFA2009}.

\subsubsection{Asymptotic Convergence} 
When $\gamma \rightarrow 0$, the convergence of FA is similar to that of PSO where the convergence was analyzed by Clerc and Kennedy in 2002 in \cite{Clerc}. When $\gamma \rightarrow \infty$, the FA may act like a random search, though its behaviour is similar to that of Simulated Annealing (SA)
because the FA's solution is perturbed or modified in the similar way as that in the SA in this limiting case. The SA was shown to be convergent under the right-cooling conditions \cite{Bertsimas}. The reduction of the roaming randomness, i.e., $\alpha_a, \alpha_b, \cdots, \alpha_z$, in the FA can be considered as a type of cooling schedule, and thus it can be expected that FA can converge in this case. 

Let us now investigate the case when $0 < \gamma <\infty$. Given a very large number of firefly population $N$, it can be assumed that $N$ is much greater than the number of local optima. The initial locations of $N$ fireflies should be uniformly distributed over the whole search space. As the iterations of Algorithm~\ref{FAoriginal} progress, i.e., $n$ increases, these initial $N$ fireflies should converge into all locally brighter ones, i.e., the local optima  including the global ones, in a stochastic manner due to the third term in \eqref{x},  \eqref{y}, and \eqref{z}. By comparing the brightest fireflies amongst the locally brighter ones, i.e., the best solutions  amongst the local optima, the global optima can be attained. Theoretically, these fireflies will reach the global optimal when $N\rightarrow \infty$ and $n\gg 1$. However, it has been reported in the related literature that the FA converges with less than 50 to 100 generations \cite{YangFA2008,YangFA2009}.

In sections~\ref{CogBeamSec}, \ref{RIS_Section}, and \ref{RIS_WPT_Sec}, we present how the proposed FA can be adopted to solve optimization problems for transmit beamforming designs.\footnote{The original FA has been discretized to solve various discrete or combinatorial optimization problems \cite{Yang_cuckoo_2014}.
For example, Osaba et al. \cite{Osaba} used a discrete FA to
solve rich vehicle routing problems.} Hereafter, ``min'' and ``s. t.'' are, respectively, used to represent ``minimize'' and ``subject to''.
\section{Transmit Beamforming}
In this section we consider a classic transmit beamforming problem with a well-known iterative method based on uplink-downlink duality. We then introduce our FA solution to the problem.

\subsection{Problem Formulation}
\subsubsection{Problem Formulation} Consider an $M_t$-antenna BS serving $U$ single-antenna mobile users.  Let $\mathbf{h}^\text{H}_{i}\in\mathbb{C}^{1 \times M_t}$, $\mathbf{w}_{i}\in\mathbb{C}^{M \times 1}$ and $s_i$, respectively, be the channel between the $i$-th user and the BS, the information-beamforming vector and the data symbol for the $i$th user. The overall signal received by the $i$th user is $y_{i}=\sum_{j=1}^{U}\mathbf{h}^\text{H}_{i}\mathbf{w}_{j}s_{j}+n_{i}$ where   $n_{i}$ is a zero mean circularly
symmetric complex Gaussian noise with variance $\sigma^2$, i.e.,
$n_{i}\sim\mathcal{CN}(0,\sigma^2)$, at the user.
Let $\mathbf{R}_{i}= \mathbf{h}_{i} \mathbf{h}^{\text{H}}_{i}$  represent the instantaneous channel state information (CSI) or $\mathbf{R}_{i}= \mathbb{E}\left(\mathbf{h}_{i}\mathbf{h}^{\text{H}}_{i}\right)$ denote the statistical CSI,  $\{\mathbf{w}_{i}\}=\{\mathbf{w}_{1},\mathbf{w}_{2},\cdots,\mathbf{w}_{U} \}$ be the set of
candidate information-beamforming vectors for all users. Assuming that $\mathbb{E}\left(|s_{i}|^2\right)=1$, the SINR at the $i$-th user is 
\begin{equation}
\textrm{SINR}_{i}=\frac{\mathbf{w}_{i}^\text{H}\mathbf{R}_{i}\mathbf{w}_{i}}
{\sum_{j=1,j \neq i}^U\mathbf{w}_{j}^\text{H}\mathbf{R}_{i}\mathbf{w}_{j}
+\sigma^2}.\label{second}
\end{equation}

We design the set of beamforming vectors $\{\mathbf{w}_{i}\}$ such that the BS's total transmit power is minimized while maintaining the SINR level at each user above the required threshold. To that end, the problem is formulated as follows: 
\begin{equation}
\begin{aligned}\label{classic_Tx_beamofming}
& \displaystyle \min_{\mathbf{w}_{i}} & &
\sum_{i=1}^U\mathbf{w}_{t}^H\mathbf{w}_{t}\\
& \text{s.\ t.}\ & &\frac{\mathbf{w}^H_{i}\mathbf{R}_{i}\mathbf{w}_{i}}
{\sum_{j =1,j \neq i}^U\mathbf{w}^H_{j}\mathbf{R}_{i}
\mathbf{w}_{j}+\sigma^2_i}\geq \gamma_{i}, \ \forall i \in \{1,\cdots,U\},
\end{aligned}
\end{equation}
where $\gamma_i$ is the required SINR level for the $i$-th user. Problem \eqref{classic_Tx_beamofming} is known as  non-convex due to the SINR constraint.

\subsubsection{Iterative Approach} An elegant approach to solve \eqref{classic_Tx_beamofming} was introduced in \cite{Farrokh} based on uplink-downlink duality where the optimal solution of the downlink problem can be sought via solving the following dual-uplink problem:\footnote{This approach was also adopted for transmit beamforing problems in coordinated multi-point (CoMP) transmissions, see e.g., \cite{Wei} and \cite{TuanTcom2013}.}
\begin{equation}
\begin{aligned}
& \displaystyle \min_{p_i} & & \sum^{U}_{i=1}p_i\\
& \text{subject\ to}\ & &\mathbf{p}\succeq \bf{\Gamma}\mathbf{t}(\mathbf{p}),\label{virtualprob2}
\end{aligned}
\end{equation}
where $\mathbf{p}=\begin{bmatrix} p_1 & p_2 & \cdots & p_U \end{bmatrix}^T$, $\mathbf{\Gamma}=\text{diag}\left[\gamma_1, \gamma_2, \cdots, \gamma_U\right]$, $\mathbf{t}(\mathbf{p})=\begin{bmatrix} t_1\left(\mathbf{p}\right) & t_2\left(\mathbf{p}\right) & \cdots & t_U\left(\mathbf{p}\right) \end{bmatrix}^T$,
\begin{equation}
t_i\left(\mathbf{p}\right)=\text{arg}\displaystyle \min_{\hat{\mathbf{w}}_i}\frac{\hat{\mathbf{w}}^{H}_{i}\mathbf{Q}_i\left(\mathbf{p}\right)\hat{\mathbf{w}}_i}{\hat{\mathbf{w}}^{H}_{i}\mathbf{R}_{i}\hat{\mathbf{w}}_i}, \label{interfere}
\end{equation}
$\mathbf{Q}_i\left(\mathbf{p}\right) =\left(\sum^{U}_{t=1,t\neq i}p_t\mathbf{R}_{t}+\sigma^2_i\mathbf{I}\right)$, $p_i=\lambda_i\sigma^2_i$ is the dual-uplink power for $i$-th user, $\lambda_i$ is the $i^{\text{th}}$ Lagrange multiplier associated with the $i^{\text{th}}$ constraint in \eqref{classic_Tx_beamofming},  and $\hat{\mathbf{w}}_i$, i.e., $\hat{\mathbf{w}}_i^H\hat{\mathbf{w}}_i=1$, is the dual-uplink beamforming vector for $i$-th user. Starting from any positive initial value of $\mathbf{p}  \left( 0 \right)$, the solution for the dual-uplink problem \eqref{virtualprob2} can be found iteratively as $\mathbf{p}\left(n+1\right)= {\bf\Gamma}\mathbf{t}\left(\mathbf{p}  \left( n \right)\right)$.  The iterative downlink algorithm to find optimal solutions for \eqref{classic_Tx_beamofming}  is summarised in algorithm \ref{TxIterative}.
\begin{algorithm}
\caption{Iterative algorithm for problem \eqref{classic_Tx_beamofming}  }
\label{TxIterative}
\begin{algorithmic}[1]
    \State \textbf{Input:} $\mathbf{\Gamma}=\text{diag}\left[\gamma_1, \gamma_2, \cdots, \gamma_U\right]$, $\mathbf{R}_i,\ \forall i$, number of iterations $T$.
    \State Initialize $\mathbf{p}  \left( 1 \right)\succeq 0$.
    \For{$n=1:T$}
    \For{$i=1:U$}
    \State Find $\hat{\mathbf{w}}_i\left( n \right)$ as the dominant eigenvector of the matrix $\mathbf{G}_i(n)=p_i\left( n \right)\mathbf{Q}_i^{-1}\left(\mathbf{p}\left(n\right)\right)\mathbf{R}_i$
    \State Calculate $t_i\left(\mathbf{p}\left( n \right)\right)=\frac{\hat{\mathbf{w}}^{H}_{i}(n)\mathbf{Q}_i\left(\mathbf{p}\left(n\right)\right)\hat{\mathbf{w}}_i(n)}{\hat{\mathbf{w}}^{H}_{i}(n)\mathbf{R}_{i}\hat{\mathbf{w}}_i(n)}$.
    \EndFor
    \State Update $\mathbf{p}\left(n+1\right)= {\bf\Gamma}\mathbf{t}\left(\mathbf{p}  \left( n \right)\right)$. 
    \EndFor
    \State $p_i^{\star}=\mathbf{p}\left(n+1\right)$ and $\hat{\mathbf{w}}_i^{\star}=\hat{\mathbf{w}}_i\left( n +1\right)$.
    \State \textbf{Output: } $\mathbf{w}_i^{\star}=\sqrt{p_i^{\star}} \hat{\mathbf{w}}_i^{\star}$.
\end{algorithmic}
\end{algorithm}
\subsection{Proposed Firefly Algorithm} We rewrite \eqref{classic_Tx_beamofming} as
\begin{equation}
\begin{aligned}\label{TxBF_rearrange}
& \displaystyle \min_{\mathbf{W}} & & f\left( \mathbf{W}\right)\\
& \text{s.\ t.}\ & &
d_i\left(\mathbf{W}\right)\leq 0 , \ \forall i,
\end{aligned}
\end{equation}
where $\mathbf{W}=\begin{bmatrix}\mathbf{w}_1, \mathbf{w}_2, \cdots, \mathbf{w}_U \end{bmatrix} \in\mathbb{C}^{M_t \times U}$, $f\left( \mathbf{W}\right)=\sum_{i=1}^U\mathbf{w}^\text{H}_{i}\mathbf{w}_{i}$, $d_i(\mathbf{W})= -\mathbf{w}_{i}^\text{H}\mathbf{R}_{i}\mathbf{w}_{i} +
   \gamma_i\sum_{j=1,j \neq i}^U\mathbf{w}_{j}^\text{H}\mathbf{R}_{i}\mathbf{w}_{j}
+\gamma_i\sigma^2_i$. Using the penalty method, we recast  \eqref{cog_rearrange} into an unconstrained problem as:
\begin{equation}
\begin{aligned}\label{sercure_wpt_unconstrained}
& \displaystyle \min_{\mathbf{W}} & & f\left( \mathbf{W}\right)+ P(\mathbf{W}),
\end{aligned}
\end{equation}
where $P(\mathbf{W})$ is the penalty term given as:
\begin{eqnarray}
    P(\mathbf{W})=\sum_{i=1}^U\lambda_i\text{max}\left\{0, d_i(\mathbf{W})\right\}^2,\label{penalty2}
\end{eqnarray}
with $\lambda_i>0$ is the penalty constant.

Let $\left\{\mathbf{W}_i\right\}=\left\{\begin{bmatrix}\mathbf{w}_1^i, \mathbf{w}_2^i, \cdots, \mathbf{w}_U^i  \end{bmatrix}\right\} $ be the $i$-th firefly. We initialize a population of $N$ fireflies $\left\{\mathbf{W}_i\right\}$, $i\in \{1,2,\cdots, N\}$, and define the light density of the firefly  $\left\{\mathbf{W}_i\right\}$ as: 
\begin{equation}
   I_i\left(\mathbf{W}_i\right)=\frac{1}{f\left( \mathbf{W}_i\right)+P(\mathbf{W}_i)}. \label{lightTxBF}
\end{equation}

For any two fireflies $i$ and $j$ in the population, if $I_j\left(\mathbf{W}_j\right)~>~I_i\left(\mathbf{W}_i\right)$ then the firefly $i$ will move toward the firefly $j$ as:
\begin{eqnarray}
    \mathbf{W}_i^{(n+1)}&=&\mathbf{W}_i^{(n)}+\beta_0 e^{-\gamma \left(r_{ij}^{(n)}\right)^2}\left(\mathbf{W}_j^{(n)}-\mathbf{W}_i^{(n)} \right)+\alpha^{(n)}\mathbf{V},\label{secureFAmoveW}
\end{eqnarray}
where $r_{ij}^{(n)}=|| (\mathbf{W}_j^{(n)}-\mathbf{W}_i^{(n)}||$ is the Cartesian distance,  $\beta_0$ is the attractiveness at $r_{ij}^{(n)}=0$, $\gamma$ presents the variation of of the attractiveness. The second term of \eqref{secureFAmoveW}  represent the attraction. The third term of \eqref{secureFAmoveW} is a randomization comprised of a randomization factor $\alpha^{(n)}$ and a matrix of random numbers  $\mathbf{V}\in\mathbb{C}^{M_t \times U}$. The random factor $\alpha^{(n)}$ and the elements of $\mathbf{V}$  are drawn from either a Gaussian or an uniform distribution.

It can be seen that problem \eqref{TxBF_rearrange} is a special case of the proposed framework \eqref{framework} where the objective and constraints are functions of optimization variable $\mathbf{W}$. Hence, the proposed FA has the same steps as those in Algorithm~\ref{FAoriginal} except  steps 3, 16, 18 and 19 given in Algorithm~\ref{TxBF_FA}.

\begin{algorithm}
\caption{Modified generalized FA for solving \eqref{TxBF_rearrange} }\label{TxBF_FA}
\begin{algorithmic}
\State \textbf{Input:} {\it FA parameters: } $N$, $T$, $\lambda_i$,  $\beta_0$; {\it Optimization data:} $\textbf{R}_{i}$, $\sigma^2_i$, $\gamma_i$; 
\State Step 3: Evaluate the light intensities of $N$ fireflies as \eqref{lightTxBF};
\State Step 16: Move firefly $i$ towards firefly $j$ as \eqref{secureFAmoveW};
\State Step 18: Attractiveness varies with distance via $e^{-\gamma \left(r_{ij}^{(n)}\right)^2}$;
\State Step 19: Evaluate new solutions; update $I_i(\mathbf{W}_i)$ as \eqref{lightTxBF};
\State \Return $\mathbf{W}^{\star}$.
\end{algorithmic}
\end{algorithm}

\subsection{Complexity Analysis} The complexity of algorithm~\ref{TxIterative} is described in the following lemma.
\begin{lemma}\label{lemIterativeTxBF}
The computational complexity of algorithm~\ref{TxIterative} is on the order of $T\left [ U(M_t^3+M_t^2+M_t\log M_t) +U\right ]$.
\end{lemma}
\begin{proof}
The proof is based on the observation that complexities of steps 5, 6 and 8 are, respectively, on the order of $M_t^3+M_t\log M_t$, $M_t^2$ and $U$. 
\end{proof}
\begin{lemma}\label{lemFATxBF}
The computational complexity of Algorithm~\ref{TxBF_FA} is on the order of:
\begin{eqnarray}
    T N^2 \left[ M_t^2+NUM_t(1+UM_t)\right]+T N \log{N}+NM_tU\nonumber \\ +NUM_t(1+UM_t)+N\log{N}.\label{Comp_FATxBF}
\end{eqnarray}
\end{lemma}
\begin{proof}
Due to space limitation, we provide main observations to derive \eqref{Comp_FATxBF} as follows. The dominant terms of the computational complexity of Algorithm~\ref{TxBF_FA} are at steps 2, 3, 4, 16, 19, and 22. The complexity of generating $N$ matrices, each matrix of size $M_t\times U$, in step 2 is on the order of $NM_tU$. The complexity of evaluating each $d_i(\mathbf{W})$ is on the order of $UM_t^2$, while the complexity of evaluating $\sum_{t=1}^U \mathbf{w}_{t}^H\mathbf{w}_{t}$  is on the order of $UM_t$.\footnote{Here, we adopt the schoolbook iterative algorithm to evaluate complexity of the multiplication of two matrices of sizes $n\times m$ and $m \times p$ as the order of $nmp$.} Hence the complexity of calculating the light density for $N$ fireflies, i.e., steps 3 and 19, is on the order of  $N(UM_t+U^2M_t^2)=NUM_t(1+UM_t)$. The complexity of ranking $N$ firefly in steps 4 and 22 is $N\log{N}$. Finally, the complexity of moving a firefly in step 16 is on the order of $M_t^2$. Assuming a worst case when step 16 is executed in every inner loop of the algorithm, after some manipulations, one can arrive at \eqref{Comp_FATxBF}. 
\end{proof}

\section{Cognitive Beamforming}\label{CogBeamSec}
\subsection{Problem Formulation}
\subsubsection{Problem Formulation}
Consider a cognitive wireless communication system consisting of an $M_t$-antenna cognitive base station (BS),
$U$ active single-antenna secondary users (SUs) and $K$ single-antenna primary users (PUs). The cognitive BS is allowed to communicate with its SUs in the same frequency band owned by the primary system if its interference imposed on each PU is less than a predefined tolerable threshold of $I_{\textrm{to},k}$. The received signal at the  $t$-th SU, $ t
 \in \{1, \cdots, U\}$, is:
\begin{eqnarray}\label{signal1}
y_{t}=\mathbf{h}^H_{s,t}\mathbf{w}_{t}s_{t}+\sum_{j=1,j
\neq t}^{U}\mathbf{h}^H_{s,t}\mathbf{w}_{j}s_{j}+n_{t},
\end{eqnarray}
where $\mathbf{h}^H_{s,t}\in\mathbb{C}^{1 \times M_t}$ is the channel coefficient
of the wireless link between the $t$-th SU and the cognitive BS; $\mathbf{w}_{t}\in\mathbb{C}^{M_t \times 1}$  and
$s_{t}\sim\mathcal{CN}(0,1)$ are, respectively, the beamforming vector and the data symbol associated to the $t$-th SU; and $n_{t}\sim\mathcal{CN}(0,\sigma^2_t)$ is a zero mean circularly
symmetric complex Gaussian noise with variance $\sigma^2_t$, at the $t$-th SU. Let $\mathbf{R}_{s,t}=\mathbb{E}\left( \mathbf{h}_{s,t}\mathbf{h}^H_{s,t} \right)$ for the statistical CSI and $\mathbf{R}_{s,t}=\mathbf{h}_{s,t}\mathbf{h}^H_{s,t}$ for the instantaneous CSI. The SINR at the $t$-th SU can be expressed as:
\begin{equation}
\text{SINR}_{t}=\frac{\mathbf{w}_{t}^H \mathbf{R}_{s,t}\mathbf{w}_{t}}{\sum_{j=1,j \neq t}^{U}\mathbf{w}_{j}^H\mathbf{R}_{s,t}\mathbf{w}_{j}+\sigma^2_t}.\label{second}
\end{equation}

Let $\mathbf{h}^H_{p,k}\in\mathbb{C}^{1 \times M_t}$ be the 
channel coefficient of the wireless link between the $k$-th PU, ${k}\in \{1, \cdots, K \}$, and the cognitive
BS, $\mathbf{R}_{p,k}=\mathbb{E}\left( \mathbf{h}_{p,k}\mathbf{h}^H_{p,k} \right)$ for the statistical CSI and $\mathbf{R}_{p,k}=\mathbf{h}_{p,k}\mathbf{h}^H_{p,k}$ for the instantaneous CSI. The
total interference power imposed on the $k$-th PU by the cognitive BS is $\sum_{j=1}^{U}
\mathbf{w}_{j}^H\mathbf{R}_{p,k}\mathbf{w}_{j}.$

Our objective is to design downlink beamforming
vectors for the SUs that minimize the cognitive BS transmit power while maintaining the required SINR level for every SU and keeping
the interference level imposed at each PU receiver below the predefined tolerable threshold. The
optimization problem to design beamforming vectors is cast as:
\begin{equation}
\begin{aligned}\label{cog_perfectCSI}
& \displaystyle \min_{\mathbf{w}_{t}} & &
\sum_{t=1}^U\mathbf{w}_{t}^H\mathbf{w}_{t}\\
& \text{s.\ t.}\ & &\frac{\mathbf{w}^H_{t}\mathbf{R}_{s,t}\mathbf{w}_{t}}
{\sum_{j =1,j \neq t}^U\mathbf{w}^H_{j}\mathbf{R}_{s,t}
\mathbf{w}_{j}+\sigma^2_t}\geq \eta_{t}, \ \forall t \in \{1,\cdots,U\},
\\ & & & \sum_{j =1}^U \mathbf{w}^H_{j}
\mathbf{R}_{p,k}\mathbf{w}_{j}\leq  I_{\textrm{to},k}, \ \forall k \in \{1,\cdots,K\},
\end{aligned}
\end{equation}
where $\eta_t$ is the required SINR level for the $t$-th SU. Due to the SINR constraint, problem \eqref{cog_perfectCSI} is non-convex.
\subsubsection{SDP Approach}
For the sake of completeness, we provide a review on a traditional approach to solve \eqref{cog_perfectCSI} using semidefinite programming (SDP). We first form a new optimization variable $\mathbf{F}_t=\mathbf{w}_t\mathbf{w}_t^H$ where $\mathbf{F}_{t}\succeq \mathbf{0}$, $\mathbf{F}_{t}\in \mathbb{H}^{M_t\times M_t}$, and $\mathbf{F}_{t}$ is a rank-one matrix.\footnote{A matrix is rank-one
if and only if it has only one linearly independent column/row.} We then utilize the identity $\mathbf{x}^H\mathbf{X}\mathbf{x}=\textrm{Tr}(\mathbf{X}\mathbf{x}\mathbf{x}^H)$ to rewrite \eqref{cog_perfectCSI} as:
\begin{equation}
\begin{aligned}\label{cog_perfect_1}
& \displaystyle \min_{\mathbf{F}_t\in \mathbb{H}^{M\times M}} & & \sum_{t=1}^U\textrm{Tr}\left(\mathbf{F}_t\right)\\
& \text{s.\ t.}\ & & \left(1+\frac{1}{\eta_t} \right) \textrm{Tr}\left(\mathbf{R}_{s,t}\mathbf{F}_t\right)-\sum_{j=1}^U\textrm{Tr}
\left(\mathbf{R}_{s,t}\mathbf{F}_{j}\right)-\sigma_t^2\geq0,\ \forall t,
\\ & & &   I_{\textrm{to},k}-\sum_{j=1}^U \textrm{Tr}\left(\mathbf{R}_{p,k}\mathbf{F}_{j}\right) \geq 0, \ \forall k,\\ &&& \mathbf{F}_t\succeq \mathbf{0},\ \forall t,
\end{aligned}
\end{equation}
where $t \in \{1,\cdots,U\}$, $k \in \{1,\cdots,K\} $. 

Problem \eqref{cog_perfect_1} is in a standard SDP form. Hence,  its optimal solution can be obtained in a polynomial time by using a general purpose IPM, e.g., CVX which is a Matlab based modeling system for constructing and solving disciplined convex programs \cite{cvx2015}. In arriving at \eqref{cog_perfect_1}, we have relaxed the rank-one constraint on $ \mathbf{F}_t, \ \forall t$. If the solution of \eqref{cog_perfect_1} does not have rank-one, then further computation resources are required to derive a sub-optimal solution via some rank-one approximations or the Gaussian randomize procedure \cite{Zhi}.
\subsection{Proposed Firefly Algorithm}
Here, we adopt the generalized FA in Algorithm~\ref{FAoriginal} to solve \eqref{cog_perfectCSI}. Rearranging the constraint, we rewrite \eqref{cog_perfectCSI} as:

\begin{equation}
\begin{aligned}\label{cog_rearrange}
& \displaystyle \min_{\mathbf{W}} & & f\left( \mathbf{W}\right)\\
& \text{s.\ t.}\ & &\
\phi_t(\mathbf{W})\leq 0 , \ \forall t \in \{1,\cdots,U\},
\\ & & &\varphi_k(\mathbf{W}) \leq  0, \ \forall k \in \{1,\cdots,K\},
\end{aligned}
\end{equation}
where $\mathbf{W}=\begin{bmatrix}\mathbf{w}_1, \mathbf{w}_2, \cdots, \mathbf{w}_U \end{bmatrix} \in\mathbb{C}^{M_t \times U}$, $f\left( \mathbf{W}\right)=\sum_{t=1}^U\mathbf{w}_{t}^H\mathbf{w}_{t}$, $\phi_t(\mathbf{W})=\eta_{t}\sum_{j =1,j \neq i}^U\mathbf{w}^H_{j}\mathbf{R}_{s,t}
\mathbf{w}_{j}+\eta_{t}\sigma^2_t -\mathbf{w}^H_{t}\mathbf{R}_{s,t}\mathbf{w}_{t}$ and $\varphi_k(\mathbf{W})=\sum_{j =1}^U \mathbf{w}^H_{j}
\mathbf{R}_{p,k}\mathbf{w}_{j}-I_{\textrm{to},k}$.
Using the penalty method, we first transform  \eqref{cog_rearrange} into an unconstrained problem as:
\begin{equation}
\begin{aligned}\label{unconstrained}
& \displaystyle \min_{\mathbf{W}} & & f\left( \mathbf{W}\right)+ P(\mathbf{W}),
\end{aligned}
\end{equation}
where $P(\mathbf{W})$ is the penalty term given as:
\begin{eqnarray}
    P(\mathbf{W})=\sum_{t=1}^U\lambda_t\text{max}\left\{0, \phi_t(\mathbf{W}) \right\}^2\label{penalty}+\sum_{k=1}^{K}\rho_k \text{max}\left\{0, \varphi_k(\mathbf{W})\right\}^2,
\end{eqnarray}
with $\lambda_t>0$ and $\rho_k>0$ are penalty constants. 

Let $\mathbf{W}_i=\begin{bmatrix}\mathbf{w}_1^i, \mathbf{w}_2^i, \cdots, \mathbf{w}_U^i  \end{bmatrix} \in\mathbb{C}^{M_t \times U}$ be the firefly $i$. We initialize a population of $N$ fireflies $\mathbf{W}_i$, $i\in \{1,2,\cdots, N\}$, and define the light density of the firefly  $\mathbf{W}_i$ as: 

\begin{equation}
   I_i\left(\mathbf{W}_i\right)=\frac{1}{f\left( \mathbf{W}_i\right)+P(\mathbf{W}_i)}. \label{lightCog}
\end{equation}

For any two fireflies $i$ and $j$ in the population, if $I_j\left(\mathbf{W}_j\right)~>~I_i\left(\mathbf{W}_i\right)$ then the firefly $i$ will move toward the firefly $j$ as:
\begin{equation}
    \mathbf{W}_i^{(n+1)}=\mathbf{W}_i^{(n)}+\beta_0 e^{-\gamma \left(r_{ij}^{(n)}\right)^2}\left(\mathbf{W}_j^{(n)}-\mathbf{W}_i^{(n)} \right)+\alpha^{(n)}\mathbf{V},\label{FAmove}
\end{equation}
where $r_{ij}^{(n)}=|| (\mathbf{W}_j^{(n)}-\mathbf{W}_i^{(n)}||$ is the Cartesian distance,  $\beta_0$ is the attractiveness at $r_{ij}^{(n)}=0$, $\gamma$ presents the variation of of the attractiveness. The second term of \eqref{FAmove} captures the attraction. The third term of \eqref{FAmove} is a randomization comprised of a randomization factor $\alpha^{(n)}$ and a matrix of random numbers  $\mathbf{V}\in\mathbb{C}^{M_t \times U}$. The random factor $\alpha^{(n)}$ and the elements of $\mathbf{V}$ are drawn from either a Gaussian or an uniform distribution. 

It can be seen that problem \eqref{cog_rearrange} is a special case of the proposed framework \eqref{framework} where the objective and constraints are functions of only one optimization variable $\mathbf{W}$. Hence, the proposed FA has the same steps as those in Algorithm~\ref{FAoriginal} except  steps 3, 16, 18 and 19 given in Algorithm~\ref{FAcognitive}.

\begin{algorithm}
\caption{Modified generalized FA for solving \eqref{cog_perfectCSI}}\label{FAcognitive}
\begin{algorithmic}
\State \textbf{Input:} {\it FA parameters: } $N$, $T$, $\lambda_t$, $\rho_k$, $\beta_0$, $\gamma$; 
{\it Optimization data:} $\textbf{R}_{s,t}$, $\textbf{R}_{p,k}$, $\sigma^2_t$, $\eta_t$, $I_{\text{to},k}$; 
\State Step 3: Evaluate the light intensities of $N$ fireflies as \eqref{lightCog};
\State Step 16: Move firefly $i$ towards firefly $j$ as \eqref{FAmove}; 
\State Step 18: Attractiveness varies with distance via $e^{-\gamma \left(r_{ij}^{(n)}\right)^2}$;
\State Step 19: Evaluate new solutions; update $I_i(\mathbf{W}_i)$ as \eqref{lightCog};
\State \Return $\mathbf{W}^{\star}$.
\end{algorithmic}
\end{algorithm}
\subsection{Complexity Analysis}
We investigate the complexity of solving \eqref{cog_perfect_1} in a worst-case runtime of the IPM  followed by the complexity analysis of the proposed FA. We start by the following definition.
\begin{definition}
At a given $\varepsilon>0$, the set of $\{\mathbf{F}_t^{\varepsilon} \}$ is an $\varepsilon$-solution to problem \eqref{cog_perfect_1}, i.e., an acceptable solution with the accuracy of $\varepsilon$, if
\begin{equation}
    \sum_{t=1}^U\textrm{Tr}\left(\mathbf{F}_t^{\varepsilon}\right)\leq \displaystyle \min_{\mathbf{F}_t\in \mathbb{H}^{M\times M}}  \sum_{t=1}^U\textrm{Tr}\left(\mathbf{F}_t\right) +\varepsilon.
\end{equation}
\end{definition}
The number of decision variables of \eqref{cog_perfect_1} is $M_t^2$. The complexity of \eqref{cog_perfect_1} is described in the following lemma.
\begin{lemma}\label{lemSDPcog}
The computational complexity to attain $\varepsilon$-solution to \eqref{cog_perfect_1} is on the order of:
\begin{align}
\ln{\left(\varepsilon^{-1}\right)}\sqrt{U(M_t+1)+K}\Big[ (M_t^2+1)(U+K)  \nonumber \\
     +UM_t^2(M_t^2+M_t)+M_t^4 \Big] M_t^2.\label{Comp_SDPcog}
\end{align}
\end{lemma}
\begin{proof}
We sketch some main steps to arrive at the lemma due to space limitation. It can be observed that \eqref{cog_perfect_1} has $(U+K)$ linear-matrix-inequality (LMI) constraints of size 1 and $U$ LMI constraints of size $M_t$. One can follow the same steps as in \cite[Section V-A]{Kun-Yu2014} to derive the following facts: (i) the iteration complexity is on the order of $\ln{\left(\varepsilon^{-1}\right)}\sqrt{U(M_t+1)+K}$, and (ii)  the per-iteration complexity is on the order of $\left[ (M_t^2+1)(U+K)+UM_t^2(M_t^2+M_t)+M_t^4\right]M_t^2$.
\end{proof}
\begin{lemma}\label{lemFAcog}
The computational complexity of Algorithm~\ref{FAcognitive} is on the order of:
\begin{eqnarray}
    T N^2 \left[ M_t^2+NUM_t(1+UM_t+KM_t)\right]+T N \log{N}+NM_tU\nonumber \\ +NUM_t(1+UM_t+KM_t)+N\log{N}.\label{Comp_FAcog}
\end{eqnarray}
\end{lemma}
\begin{proof}
Due to space limitation, we provide main observations to derive \eqref{Comp_FAcog} as follows. The dominant terms of the computational complexity of Algorithm~\ref{FAcognitive} are at steps 2, 3, 4, 16, 19, and 22. The complexity of generating $N$ matrices, each matrix of size $M_t\times U$, in step 2 is on the order of $NM_tU$. The complexity of evaluating each $\phi_t(\mathbf{W})$ or $\varphi_k(\mathbf{W})$ is on the order of $UM_t^2$, while the complexity of evaluating $\sum_{t=1}^U \mathbf{w}_{t}^H\mathbf{w}_{t}$  is on the order of $UM_t$. Hence the complexity of calculating the light density for $N$ fireflies, i.e., steps 3 and 19, is on the order of  $N(UM_t+U^2M_t^2+KUM_t^2)=NUM_t(1+UM_t+KM_t)$. The complexity of ranking $N$ firefly in steps 4 and 22 is $N\log{N}$. Finally, the complexity of moving a firefly in step 16 is on the order of $M_t^2$. Assuming a worst case when step 16 is executed in every inner loop of the algorithm, after some manipulations, one can arrive at \eqref{Comp_FAcog}. 
\end{proof}

\section{Reconfigurable Intelligent Surface-Aided Beamforming}\label{RIS_Section}
\subsection{Problem Formulation}\label{ris_problem_formulation}
\subsubsection{Problem Formulation}
Consider a communication system comprising of an $M_t$-antenna BS communicating  with $U$ single-antenna mobile users in which the direct communication links between the BS and its mobile users are blocked, e.g., because of high building etc., \cite{Tuan2021}. To circumvent the problem, an $N_t$-reflective-element RIS is utilized to support the communication. Let $\mathbf{H} = [\mathbf{h}_1, \ldots, \mathbf{h}_{N_t}] \in \mathbb{C}^{M_t \times N_t}$ represent the channel coefficients between the BS and the RIS and $\mathbf{g}_i =[g_{i1}, \ldots, g_{iN_t}]^T \in \mathbb{C}^{N_t \times 1}$ be the channel coefficients between the RIS and the $i$-th user.

Let $x_i$, i.e., $\mathbb{E}[|x_i|^2]=1$, and $\mathbf{w}_i \in \mathbb{C}^{M_t\times 1}$, respectively, represent the data symbol and the active beamforming vector for the $i$-th user. Each reflective element  of the RIS generates a phase shift to support the communication between the BS and the mobile users. Let $\theta_k$ be the phase shift at the $k$-th reflective element and let $\pmb{\theta} =[\theta_1, \ \theta_2, \ \cdots, \ \theta_{N_t}]^T$ denote the phase-shift coefficients generated by the RIS with $|\theta_k| \leq 1$ and $\textrm{arg}(\theta_k)\in [-\pi,\pi), \forall k = 1, \ldots, N_t$. Vector $\pmb{\theta}$ is the passive beamforming vector for the RIS. The signal arrived at the~$i$-th user is:
\begin{eqnarray}\label{re01}
y_i&=&\mathbf{g}_i^H \textrm{diag}(\pmb{\theta})^H \mathbf{H}^H \mathbf{w}_{i} x_{i} +\mathbf{g}_i^H \textrm{diag}(\pmb{\theta})^H \mathbf{H}^H \sum_{j=1,j \neq i}^{U}\mathbf{w}_{j} x_{j}+ n_i,\nonumber\\
&=&\pmb{\theta}^H \mathbf{G}_i^H \mathbf{w}_{i} x_{i} +\pmb{\theta}^H \mathbf{G}_i^H \sum_{j=1,j\neq i}^{U}\mathbf{w}_{j} x_{j} + n_i,
\end{eqnarray}
where $\mathbf{G}_i^H=\textrm{diag}(\mathbf{g}_i^{\ast})\mathbf{H}^H \in \mathbb{C}^{N_t \times M_t}$ and $n_i\sim \mathcal{CN}(0,\sigma^2)$ represents the additive noise measured at the $i$-th user. Furthermore, let $\{\mathbf{w}_{i}\}=\{\mathbf{w}_1, \mathbf{w}_2,\cdots, \mathbf{w}_U\}$ denote the set of active beamforming vectors, and  $\textrm{SINR}_i( \{\mathbf{w}_{i}\},\pmb{\theta})$ be the SINR at the $i$-th user. One can write:
\begin{equation}\label{sinr01}
\textrm{SINR}_i\left( \{\mathbf{w}_{i}\},\pmb{\theta}\right) =\frac{|\pmb{\theta}^H \mathbf{G}_i^H \mathbf{w}_i|^2}{ \sum\limits_{j=1,j\neq i}^{U}|\pmb{\theta}^H \mathbf{G}_i^H\mathbf{w}_{j}|^2+\sigma_i^2}.
\end{equation}
The optimization is posed as follows:
\begin{equation} \label{IRS_PerfectCSI}
\begin{aligned}
& \underset{ \{\mathbf{w}_{i} \},\ \pmb{\theta} }{\textrm{min}} & &
\sum_{i=1}^U   \mathbf{w}_{i}^H\mathbf{w}_i\\
& \mbox{s.\ t.}\ & & \textrm{SINR}_{i}\left( \{\mathbf{w}_{i}\},\pmb{\theta}\right) \geq \eta_i, \forall i, \\
&&& |\theta_k| \leq 1, \forall k,
\end{aligned}
\end{equation}
where $\eta_i$ is the required SINR level measured at the~$i$-th user. Since the SINR constraint is a function of two optimization variables $\mathbf{w}_{i} $ and $\pmb{\theta}$, problem \eqref{IRS_PerfectCSI} is non-convex.
\subsubsection{Alternative Optimization Approach}
For the sake of completeness, the widely-adopted AO approach \cite{Tuan2021,TrungRIS2020,NamRIS2021,Peng2022} is represented here as a baseline to solve \eqref{IRS_PerfectCSI}. 
Let $\mathbf{F}_i=\mathbf{w}_i\mathbf{w}_i^H$, and $\boldsymbol\Theta=\pmb{\theta}\pmb{\theta}^H$, i.e., $\mathrm{rank}(\mathbf{F}_i) = 1$ and $\mathrm{rank}(\mathbf{\Theta}) = 1$. 
As  $\mathbf{F}_i$ and $\boldsymbol\Theta$ are two independent variables, they can be alternatively solved \cite{Tuan2021,TrungRIS2020,NamRIS2021,Peng2022}. To that end, relaxing the rank-one constraint on $\mathbf{F}_i$ and beginning with any initial value of the reflecting coefficient matrix $\boldsymbol\Theta^{(0)}$, the following sub-problem will be solved at the $p$-th iteration:
\begin{equation}
\begin{aligned}\label{prob_IRS_W_relax}
& \underset{ \{\mathbf{F}_{i} \}}{\textrm{min}}  & & \textrm{Tr}\left(\sum_{i=1}^U \mathbf{F}_{i}\right)\\
& \mbox{s. \ t.}\ & &\textrm{Tr}\frac{\mathbf{G}_{i}\boldsymbol\Theta^{(p-1)}\mathbf{G}_i^H\mathbf{F}_{i}}{\eta_i\sigma_i^2}-\sum_{j=1,j\neq i}^U\textrm{Tr}
\frac{\mathbf{G}_{i}\boldsymbol\Theta^{(p-1)}\mathbf{G}_i^H\mathbf{F}_{j}}{\sigma_i^2}-1\geq0, \forall i,\\ 
&&& \mathbf{F}_i  \succeq \mathbf{0}, \ \forall i  \in \{1,\cdots,U\}.
\end{aligned}
\end{equation}

The reflecting coefficients $\boldsymbol\Theta^{(p)}$ is then updated from the optimal solution of \eqref{prob_IRS_W_relax} at $p$-th iteration, i.e., $\{\mathbf{F}_i^{(p)}\}$, by solving the following sub-problem \cite{Tuan2021}:
\begin{equation}
\begin{aligned}\label{prob_IRS_Theta_relax}
& \underset{ \mathbf{\Theta}}{\textrm{min}}  & & \textrm{Tr}\left( \boldsymbol\Theta\right)\\
& \mbox{s. \ t.}\ & &\textrm{Tr}\frac{\boldsymbol\Theta\mathbf{G}_i^H\mathbf{F}_{i}^{(p)}\mathbf{G}_{i}}{\eta_i\sigma_i^2}-\sum_{j=1,j\neq i}^U\textrm{Tr}
\frac{\boldsymbol\Theta\mathbf{G}_i^H\mathbf{F}_{j}^{(p)}\mathbf{G}_{i}}{\sigma_i^2}-1\geq0,  \forall i,\\ 
&&& \textrm{diag}\left(\textrm{diag}\left(\mathbf{\Theta}\right)\right)   \preceq \mathbf{I}_{N_t}, \\
&&&\mathbf{\Theta} \succeq \mathbf{0}.
\end{aligned}
\end{equation}

The AO approach repetitively solves  two SDPs \eqref{prob_IRS_W_relax} and \eqref{prob_IRS_Theta_relax} in $n_0$ iterations to obtain the solution for \eqref{IRS_PerfectCSI}. 
\begin{remark}
It is worth noticing that the AO approach approximates the originally non-convex optimization \eqref{IRS_PerfectCSI} by two sub-problems \eqref{prob_IRS_W_relax}  and \eqref{prob_IRS_Theta_relax}. Although \eqref{prob_IRS_W_relax}  and \eqref{prob_IRS_Theta_relax} are convex, the solutions to these sub-problems can be regarded as the upper bounds of the original problem \eqref{IRS_PerfectCSI} as these solutions may not be the global solution. Furthermore, the AO approach adopts the so-called semidefinite relaxation technique \cite{Mats} in which the rank-one constraints on $\mathbf{F}_i$ and $\boldsymbol\Theta$ are relaxed. If solving \eqref{prob_IRS_W_relax}  and/or \eqref{prob_IRS_Theta_relax} does not return rank-one matrices $\mathbf{F}_i$ and/or $\boldsymbol\Theta$, then a rank-one approximation or a Gaussian randomize procedure \cite{Zhi} is required to extract approximated rank-one solutions. Extracting the approximated solutions requires further computational resources yet only results in sub-optimal solutions.
\end{remark}

Motivated by the above observations, we introduce a novel FA approach to simultaneously solve $\mathbf{w}_{i} $ and $\pmb{\theta}$ for the original problem \eqref{IRS_PerfectCSI} in the following section.
\subsection{Proposed Firefly Algorithm}\label{RIS_WPT_section}
The optimization \eqref{IRS_PerfectCSI} can be expressed as
\begin{equation} \label{IRS_PerfectCSI_n}
\begin{aligned}
& \underset{ \{\mathbf{W}, \ \pmb{\theta}\} }{\textrm{min}} & &f\left( \mathbf{W}\right)
\\
& \mbox{s.\ t.}\ & & \phi_i\left( \{\mathbf{W},\pmb{\theta}\}\right) \leq 0, \forall i, \\
&&& \varphi_k\left(\theta_k\right) \leq 0, \forall k,
\end{aligned}
\end{equation}
where $\mathbf{W}=\begin{bmatrix}\mathbf{w}_1, \mathbf{w}_2, \cdots, \mathbf{w}_U \end{bmatrix} \in\mathbb{C}^{M_t \times U}$, $f\left( \mathbf{W}\right)=\sum_{i=1}^U  \mathbf{w}_{i}^H\mathbf{w}_i$,
\begin{eqnarray}
    \phi_i\left( \mathbf{W},\pmb{\theta}\right)=\eta_i\frac{\sum_{j=1}^{U}\mathbf{w}_j^H \mathbf{G}_i\pmb{\theta}\pmb{\theta}^H \mathbf{G}_i^H\mathbf{w}_j}{\sigma^2_i}+\eta_i\nonumber \\-\left( 1+\eta_i\right)\frac{\mathbf{w}_i^H \mathbf{G}_i\pmb{\theta}\pmb{\theta}^H \mathbf{G}_i^H\mathbf{w}_i}{\sigma^2_i},
\end{eqnarray}
and $\varphi_k\left(\theta_k\right)=|\theta_k|-1$.
Adopting the penalty method, \eqref{IRS_PerfectCSI_n} can be written as:
\begin{equation} \label{IRS_PerfectCSI_n2}
\begin{aligned}
& \underset{ \{\mathbf{W}, \ \pmb{\theta}\}  }{\textrm{min}} & &
f\left( \mathbf{W}\right)+ P(\mathbf{W},\pmb{\theta}),
\end{aligned}
\end{equation}
where $P(\mathbf{W},\pmb{\theta})$ is the penalty term given as:
\begin{eqnarray}
    P(\mathbf{W},\pmb{\theta})=\sum_{i=1}^U\lambda_i\text{max}\left\{0, \phi_i(\{\mathbf{W},\pmb{\theta}\}) \right\}^2\label{penalty}+\sum_{k=1}^{N_t}\rho_k \text{max}\left\{0, \varphi_k(\theta_k)\right\}^2,
\end{eqnarray}
with $\lambda_i>0$ and $\rho_k>0$ are penalty constants. 

Let $\{\mathbf{W}_t,\pmb{\theta}_t\}=\{\begin{bmatrix}\mathbf{w}_1^t, \mathbf{w}_2^t, \cdots, \mathbf{w}_U^t  \end{bmatrix}, \pmb{\theta}_t\} $ be the firefly $t$. We initialize a population of $N$ fireflies $\{\mathbf{W}_t,\pmb{\theta}_t\}$, $t\in \{1,2,\cdots, N\}$ and define the light density, i.e., the brightness, of the firefly $t$  $\{\mathbf{W}_t,\pmb{\theta}_t\}$ as: 

\begin{equation}
   I_t\left(\mathbf{W}_t,\pmb{\theta}_t\right)=\frac{1}{f\left( \mathbf{W}_t\right)+P(\mathbf{W}_t,\pmb{\theta}_t)}. \label{lightRIS}
\end{equation}

For any fireflies $t$ and $l$ amongst the population, if $I_t\left(\mathbf{W}_t,\pmb{\theta}_t\right) > I_l\left(\mathbf{W}_l,\pmb{\theta}_l\right)$ then the firefly $l$ will move toward the firefly $t$ as:
\begin{eqnarray}
    \mathbf{W}_l^{(n+1)}&=&\mathbf{W}_l^{(n)}+\beta_0 e^{-\gamma \left(r_{w,tl}^{(n)}\right)^2}\left(\mathbf{W}_t^{(n)}-\mathbf{W}_l^{(n)} \right)+\alpha^{(n)}\mathbf{V},\label{FAmoveW}\\
    \pmb{\theta}_l^{(n+1)}&=&\pmb{\theta}_l^{(n)}+\beta_0 e^{-\gamma \left(r_{\theta,tl}^{(n)}\right)^2}\left(\pmb{\theta}_t^{(n)}-\pmb{\theta}_l^{(n)} \right)+\alpha^{(n)}\mathbf{v},\label{FAmoveTheta}
\end{eqnarray}
where $r_{w,tl}^{(n)}=|| (\mathbf{W}_t^{(n)}-\mathbf{W}_l^{(n)}||$ and $r_{\theta,tl}^{(n)}=|| (\pmb{\theta}_t^{(n)}-\pmb{\theta}_l^{(n)}||$ are the Cartesian distances,  $\beta_0$ is the attractiveness at $r_{w,tl}^{(n)}=0$ and $r_{\theta,tl}^{(n)}=0$, $\gamma$ presents the variation of of the attractiveness. The second terms of \eqref{FAmoveW} and \eqref{FAmoveTheta} capture the attractions while the third terms of \eqref{FAmoveW} and \eqref{FAmoveTheta} are  randomization comprised of randomization factor $\alpha^{(n)}$, $\mathbf{V} \in\mathbb{C}^{M_t \times U} $ and   $\mathbf{v} \in\mathbb{C}^{M_t \times 1}$. The factor $\alpha^{(n)}$, the elements of $\mathbf{V} $ and   $\mathbf{v} $ are drawn from either an uniform or a Gaussian distribution. 

It can be observed that problem \eqref{IRS_PerfectCSI_n} is a special case of the proposed framework \eqref{framework} where the objective and constraints are functions of optimization variables $\mathbf{W}$ and $\pmb{\theta}$. The proposed FA for RIS has the same steps as those in Algorithm~\ref{FAoriginal} except steps 3, 16, 18 and 19 given in Algorithm~\ref{FARIS}.

\begin{algorithm}
\caption{Modified generalized FA for solving \eqref{IRS_PerfectCSI}}\label{FARIS}
\begin{algorithmic}
\State \textbf{Input:} {\it FA parameters: } $N$, $T$, $\lambda_i$, $\rho_n$, $\beta_0$; $\gamma$; 
{\it Optimization data:} $\textbf{H}$, $\textbf{g}_i$, $\sigma^2_i$, $\eta_i$, $I_{\text{to}}$; 
\State Step 3: Evaluate the light intensities of $N$ fireflies as \eqref{lightRIS};
\State Step 16: Move firefly $i$ towards firefly $j$ as \eqref{FAmoveW} and \eqref{FAmoveTheta};
\State Step 18: Attractiveness varies with distances via $e^{-\gamma \left(r_{w,ji}^{(n)}\right)^2}$ and $e^{-\gamma \left(r_{\theta,ji}^{(n)}\right)^2}$;
\State Step 19: Evaluate new solutions; update $I_i\left(\mathbf{W}_i,\pmb{\theta}_i\right)$ as \eqref{lightRIS};
\State \Return $\mathbf{W}^{\star}, \pmb{\theta}^{\star}$.
\end{algorithmic}
\end{algorithm}
\subsection{Complexity Analysis}
Here, we analyze the computational complexities of the AO and the proposed FA  for RIS-aided beamforming problem.
\begin{lemma}\label{lemAO}
The complexity of the AO approach is on the order of:
\begin{equation}
    n_o\left( \tau_1+ \tau_2\right),\label{ComIMPRIS}
\end{equation}
where 
\begin{eqnarray}
    \tau_1&=&\ln{\left(\varepsilon^{-1}\right)}\sqrt{U(M_t+1)}\Big[ (M_t^2+1)U
    +UM_t^2(M_t^2+M_t)\nonumber \\&{}&+M_t^4\Big]M_t^2, \label{ComW}\\
    \tau_2&=&\ln{\left(\varepsilon^{-1}\right)}\sqrt{U+2N_t}\left[ (N_t^2+1)(U+2N_t^2)+N_t^4\right]N_t^2.\label{ComP}
\end{eqnarray}
\end{lemma}
\begin{proof} We first give some hints to derive the computational complexity of obtaining optimal solution to problems \eqref{prob_IRS_W_relax} and \eqref{prob_IRS_Theta_relax}. With the observation that \eqref{prob_IRS_W_relax} has $U$ LMI constraints of size 1 and $U$ LMI constraints of size $M_t$, one can follow the same steps as in \cite[Section V-A]{Kun-Yu2014} to derive the complexity of solving \eqref{prob_IRS_W_relax} as $\tau_1$ given in \eqref{ComW}.

At a given $\varepsilon>0$, $\mathbf{\Theta}^{\varepsilon}$  is called an $\varepsilon$-solution to problem \eqref{prob_IRS_Theta_relax} if $\textrm{Tr}\left( \mathbf{\Theta}^{\varepsilon}\right)\leq \underset{ \mathbf{\Theta}}{\textrm{min}} \textrm{Tr}\left( \boldsymbol\Theta\right) +\varepsilon$. The number of decision variables of \eqref{prob_IRS_Theta_relax} is $N_t^2$. Observing that \eqref{prob_IRS_Theta_relax} has $U$ linear-matrix-inequality (LMI) constraints of size 1 and $2$ LMI constraints of size $N_t$, one can derive the computational complexity to attain $\varepsilon$-solution to \eqref{prob_IRS_Theta_relax} as the order of $\tau_2$ given in \eqref{ComP}.

Since the AO approach iteratively solves \eqref{prob_IRS_W_relax} and \eqref{prob_IRS_Theta_relax} in $n_o$ iterations, the complexity of AO approach is on the order of $n_o\left( \tau_1+ \tau_2\right)$.
\end{proof}
\begin{lemma}\label{lem_RIS}
The computational complexity of Algorithm~\ref{FARIS} is on the order of
\begin{eqnarray}
    &{}&T N^2 \left[ M_t^2+N_t+N\left(UM_t+U(N_t^2+M_tN_t)+N_t\right)\right]\nonumber\\
    &{}&+T N \log{N}+NM_tU+N_t N +N\log{N}\nonumber\\
    &{}&+N\left(UM_t+U(N_t^2+M_tN_t)+N_t\right).\label{Comp_FARIS}
\end{eqnarray}
\end{lemma}
\begin{proof}
The proof is based on the following observations. The dominant terms of the computational complexity of Algorithm~\ref{FARIS} are at steps 2, 3, 4, 16, 19, and 22. The complexity of generating $N$ fireflies in step 2 is on the order of $NM_tU+N_t N$. The complexities of evaluating  $\phi_i(\mathbf{W},\pmb{\theta})$, $\varphi_k(\theta_k)$, and $\sum_{i=1}^U \mathbf{w}_{i}^H\mathbf{w}_i$ are, respectively, on the order of $U(N_t^2+M_tN_t)$, $N_t$, and $UM_t$. Hence, the complexity of calculating the light density for $N$ fireflies, i.e., steps 3 and 19, is on the order of  $N\left(UM_t+U(N_t^2+M_tN_t)+N_t\right)$. The complexity of ranking $N$ firefly in steps 4 and 22 is $N\log{N}$. Finally, the complexity of moving a firefly in step 16 is on the order of $M_t^2+N_t$. Assuming a worst case when step 16 is executed in every inner loop of the algorithm, after some manipulations, one can arrive at \eqref{Comp_FARIS}. 
\end{proof}

\section{RIS-Aided Wireless Power Transfer}\label{RIS_WPT_Sec}
\subsection{Problem Formulation}
\subsubsection{Problem Formulation}
Consider a similar communication system in \ref{ris_problem_formulation}, however, the users are energy harvesting receivers (EHRs) instead of information decoding receivers. Using the same notations as in \ref{ris_problem_formulation}, the power arrived at the~$i$-th user is:
\begin{eqnarray}\label{ris_wpt}
E_i&=&\Big| \mathbf{g}_i^H \textrm{diag}(\pmb{\theta})^H \mathbf{H}^H \sum_{j=1}^{U}\mathbf{w}_{j}\Big|^2=\sum_{j=1}^{U}\mathbf{w}_{j}^H\mathbf{G}_i \pmb{\theta} \ \pmb{\theta}^H \mathbf{G}_i^H \mathbf{w}_{j}  ,
\end{eqnarray}
where $\mathbf{w}_{j}$ is the active energy beamforming vector for the $j$-th user. we interested in maximizing a total weighted sum power received at the EHRs obtained via the following optimization problem:
\begin{equation} \label{IRS_WPT}
\begin{aligned}
& \underset{ \{\mathbf{w}_{i} \},\ \pmb{\theta} }{\textrm{max}} & &
\sum_{i=1}^U   \sum_{j=1}^{U}\alpha_i\mathbf{w}_{j}^H\mathbf{G}_i \pmb{\theta} \ \pmb{\theta}^H \mathbf{G}_i^H \mathbf{w}_{j}\\
& \mbox{s.\ t.}\ & & \sum_{j=1}^{U}\mathbf{w}_{j}^H \mathbf{w}_{j} \leq P, \ |\theta_k| = 1, \forall k,
\end{aligned}
\end{equation}
where $P$ is the maximum transmit power of the BS and $\alpha_i\geq 0$ is the weighting factor for the $i$-th EHR. 
\subsubsection{Successive Convex Approximation} According to \cite{Wu2020}, for any fix $\pmb{\theta}$, only one common energy beam is sufficient. Using a successive convex approximation (SCA) technique, \cite{Wu2020} proposed an iterative algorithm to find optimal active and passive beamforming vectors for problem \eqref{IRS_WPT} as follows. Starting with an initialized value $\pmb{\theta}^{(0)}$, the optimal active beamforming vector at the $l$-th iterations is calculated as $\mathbf{w}^{(l)}=\sqrt{P}\text{eig}_{max}\left( \sum_{i=1}^U   \alpha_i\mathbf{G}_i \pmb{\theta}^{(l-1)}\pmb{\theta}^{(l-1)H} \mathbf{G}_i^H \right)$ where $\text{eig}_{max}\left( \mathbf{X}\right)$ is the maximum eigenvalue of matrix $\mathbf{X}$. The $k$-th coefficient of the RIS's phase shift vector at the $l$-th iterations is calculated as $\left[ \pmb{\theta}^{(l)}\right]_k=1 $ if $\mu_k=0$ and $\left [ \pmb{\theta}^{(l)}\right ]_k =\frac{\mu_k}{|\mu_k|}$ if $\mu_k \neq 0$, where $\mu_k = \left[ \sum_{i=1}^{U}\alpha_i \mathbf{G}_i^H \mathbf{w}^{(l)} \mathbf{w}^{(l)H}\mathbf{G}_i \pmb{\theta}^{(l-1)}\right]_k$. 
\subsection{Proposed Firefly Algorithm}
The optimization \eqref{IRS_WPT} can be expressed as
\begin{equation} \label{IRS_WPT_n}
\begin{aligned}
& \underset{ \{\mathbf{W}, \ \pmb{\theta}\} }{\textrm{min}} & &-f\left( \mathbf{W},\pmb{\theta}\right)
\\
& \mbox{s.\ t.}\ & & \phi\left( \{\mathbf{W},\pmb{\theta}\}\right) \leq 0, \\
&&& \varphi_k\left(\theta_k\right) = 0, \forall k,
\end{aligned}
\end{equation}
where $\mathbf{W}=\begin{bmatrix}\mathbf{w}_1, \mathbf{w}_2, \cdots, \mathbf{w}_U \end{bmatrix} \in\mathbb{C}^{M_t \times U}$, $f\left( \mathbf{W},\pmb{\theta}\right)=\sum_{i=1}^U   \sum_{j=1}^{U}\alpha_i\mathbf{w}_{j}^H\mathbf{G}_i \pmb{\theta}\pmb{\theta}^H \mathbf{G}_i^H \mathbf{w}_{j}$,
$\phi\left( \mathbf{W},\pmb{\theta}\right)=\sum_{j=1}^{U}\mathbf{w}_{j}^H \mathbf{w}_{j}-P$, 
and $\varphi_k\left(\theta_k\right)=|\theta_k|-1$. Adopting the penalty method, \eqref{IRS_PerfectCSI_n} can be written as:
\begin{equation} \label{IRS_PerfectCSI_n2}
\begin{aligned}
& \underset{ \{\mathbf{W}, \ \pmb{\theta}\}  }{\textrm{min}} & &
-f\left( \mathbf{W},\pmb{\theta}\right)+ P(\mathbf{W},\pmb{\theta})
\end{aligned}
\end{equation}
where $P(\mathbf{W},\pmb{\theta})=\lambda\text{max}\left\{0, \phi(\{\mathbf{W},\pmb{\theta}\}) \right\}^2\label{penalty}+\sum_{k=1}^{N_t}\rho_k \left\{\varphi_k(\theta_k)\right\}^2$, with $\lambda>0$ and $\rho_k>0$ are penalty constants. 

Let $\{\mathbf{W}_t,\pmb{\theta}_t\}=\{\begin{bmatrix}\mathbf{w}_1^t, \mathbf{w}_2^t, \cdots, \mathbf{w}_U^t  \end{bmatrix}, \pmb{\theta}_t\} $ be the firefly $t$. We initialize a population of $N$ fireflies $\{\mathbf{W}_t,\pmb{\theta}_t\}$, $t\in \{1,2,\cdots, N\}$ and define the light density, i.e., the brightness, of the firefly $t$  $\{\mathbf{W}_t,\pmb{\theta}_t\}$ as: 

\begin{equation}
  I_t\left(\mathbf{W}_t,\pmb{\theta}_t\right)=\frac{1}{-f\left( \mathbf{W}_t\right)+P(\mathbf{W}_t,\pmb{\theta}_t)}. \label{lightRIS_WPT}
\end{equation}

It can be observed that problem \eqref{IRS_WPT_n} is a special case of the proposed framework \eqref{framework} where the objective and constraints are functions of optimization variables $\mathbf{W}$ and $\pmb{\theta}$. Utilizing the firefly movements define in \eqref{FAmoveW} and \eqref{FAmoveTheta} in Section\ref{RIS_WPT_section}, the proposed FA for RIS has the same steps as those in Algorithm~\ref{FAoriginal} except steps 3, 16, 18 and 19 given in Algorithm~\ref{FARIS_WPT}.

\begin{algorithm}
\caption{Modified generalized FA for solving \eqref{IRS_WPT}}
 \label{FARIS_WPT}
\begin{algorithmic}
\State \textbf{Input:} {\it FA parameters: } $N$, $T$, $\lambda$, $\rho_k$, $\beta_0$; $\gamma$; 
{\it Optimization data:} $\textbf{H}$, $\textbf{g}_i$, $\alpha_i$,  $P$; 
\State Step 3: Evaluate the light intensities of $N$ fireflies as \eqref{lightRIS_WPT};
\State Step 16: Move firefly $i$ towards firefly $j$ as \eqref{FAmoveW} and \eqref{FAmoveTheta};
\State Step 18: Attractiveness varies with distances via $e^{-\gamma \left(r_{w,ji}^{(n)}\right)^2}$ and $e^{-\gamma \left(r_{\theta,ji}^{(n)}\right)^2}$;
\State Step 19: Evaluate new solutions; update $I_i\left(\mathbf{W}_i,\pmb{\theta}_i\right)$ as \eqref{lightRIS_WPT};
\State \Return $\mathbf{W}^{\star}, \pmb{\theta}^{\star}$.
\end{algorithmic}
\end{algorithm}
\subsection{Complexity Analysis}
Here, we analyze the complexities of the SCA approach and the proposed FA for the RIS-aided WPT beamforming. We start by introducing the following lemma.
\begin{lemma}\label{lem_SCA}
The complexity of the SCA approach is on the order of:
\begin{equation}
    m_0\left( UM_t\left(M_t+N_t \right)+M_t^3+M_t\log M_t +N_t^3+N_t^2M_t\right),\label{ComSCA}
\end{equation}
where $m_0$ is the number of iterations of the SCA approach.
\end{lemma}
\begin{proof}
    At each iteration, the complexity of evaluating $\alpha_i\mathbf{G}_i \pmb{\theta}^{(l-1)}\pmb{\theta}^{(l-1)H} \mathbf{G}_i^H $ is on the order of $U\left(M_t^2+M_tN_t \right)$. The complexities of finding a maximum eigenvalue of the $M_t \times M_t$ matrix $\alpha_i\mathbf{G}_i \pmb{\theta}^{(l-1)}\pmb{\theta}^{(l-1)H} \mathbf{G}_i^H $ based on the SVD method is on the order of $M_t^3+M_t\log M_t$. Hence, the complexity of finding $\mathbf{w}^{(l)}$ is on the order of $UM_t(M_t+N_t)+M_t^3+M_t\log M_t$. Furthermore, the complexity of calculating $\mu_k$ is on the order of $N_t^2+M_tN_t$. Therefore, the complexity of finding $\pmb{\theta}^{(l)}$ is on the order of $N_t\left( N_t^2+M_tN_t\right)$. Consequently, $m_0$ iterations of evaluating $\mathbf{w}^{(l)}$ and $\pmb{\theta}^{(l)}$ lead to \eqref{ComSCA}.
\end{proof}
\begin{lemma}\label{lem_FARIS_WPT}
The complexity of the Algorithm~\ref{FARIS_WPT} is on the order of:
\begin{eqnarray}
    &{}&T N^2 \left[ M_t^2+N_t+N\left(UM_t+U(N_t^2+M_tN_t)+N_t\right)\right]\nonumber\\
    &{}&+T N \log{N}+NM_tU+N_t N +N\log{N}\nonumber\\
    &{}&+N\left(UM_t+U(N_t^2+M_tN_t)+N_t\right).\label{Comp_WPT}
\end{eqnarray}
\end{lemma}
\begin{proof}
    Noticing that the complexities of evaluating  $\phi(\mathbf{W},\pmb{\theta})$, $\varphi_k(\theta_k)$, and $f(\mathbf{W},\pmb{\theta})$ are, respectively, on the order of $UM_t$, $N_t$, and $U\left(N_tM_t+N_t^2\right)$. One can easily show that the complexity of the Algorithm~\ref{FARIS_WPT} is the same as that of the Algorithm~\ref{FARIS}.
\end{proof}

\section{Numerical Results}
In this section, we perform simulations to evaluate the performances of the proposed FA approaches, i.e., FA approaches for transmit beamforming, cognitive cognitive beamforming, RIS-aided transmit beamforming, and RIS-aided WPT, and compare them with their iterative, SDP, and SCA counterparts.

CVX package \cite{cvx2015} is utilized to obtain the solution for the cognitive SPD approach, i.e., problem \eqref{cog_perfect_1}, and the AO approach for the RIS-aided transmit beamforming. In the AO approach, two SDPs \eqref{prob_IRS_W_relax} and \eqref{prob_IRS_Theta_relax} are alternatively solved in $n_0=10$ iterations. The setup parameters for FAs are as follows. The variation of the attractiveness $\gamma$ is set at 1. The penalty constants are set equal but they dynamically vary as $\lambda_i=\rho_k= n^2, \ \forall i, k$ where $n$ is the generation index in Algorithm~\ref{FAoriginal}. The attractiveness at zero distance is $\beta_0=1$. Finally, the initial randomization factor is $\alpha^{(0)}=0.9$  and its value at the $n$-th generation is $\alpha^{(n)}=\alpha^{(0)}0.9^n$.
\subsection{Evaluation on Transmit Beamforming} We simulate a scenario of two users, i.e., $U=2$, randomly distributed within $2$ km from their BS. The array antenna gain at the BS is 15dBi.  The noise power spectral density, noise figure at each user and the subcarrier bandwidth are, respectively, $-174$ dBm/Hz, $5$ dB and $15$ kHz wide. The path loss model is $35+34.5\log10(l)$, where $l$ is in kilometers. A log-normal shadowing with
a standard deviation of 8 dB is assumed. Furthermore, a complex Gaussian distribution is set 
with the variance of $1/2$ on each of its real and imaginary components for the downlink channel fading coefficients. Monte Carlo simulations have been carried out over 1000 channel realizations. 
\begin{figure}[t]
\centering
   \includegraphics[width=.5\textwidth]{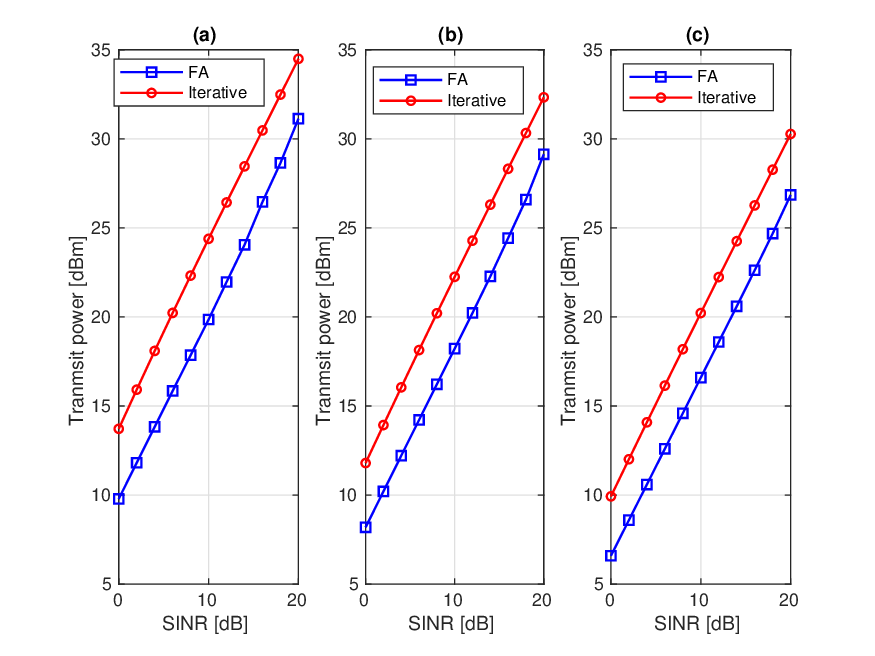}
\caption{ The total BS's transmit power versus the required SINR level with different numbers of BS's antennas: (a) 4 antennas; (b) 6 antennas; (c) 8 antennas. The firefly population is $N=30$. The number of maximum generations $T=30$.}\label{FATxBF}
\end{figure}

Fig.~\ref{FATxBF} illustrates the total transmit power of the proposed FA approach and its iterative counterpart versus the required SINR level with different numbers of BS's antennas. The results on Fig.~\ref{FATxBF} clearly show that the proposed FA approach outperforms the iterative method in obtaining lower required transmit power, i.e., around $3$ to $4$ dB lower, for all simulated setups. The results in Fig.~\ref{FATxBF} confirm the ability of the proposed FA in handling highly nonlinear and multimodal optimization problems. This power saving gain, however, comes at the price of a higher complexity. Using the parameter setup for Fig.~~\ref{FATxBF} in Lemmas  \ref{lemIterativeTxBF} and \ref{lemFATxBF}, i.e., $U=2$, $T=N=30$, $M_t=4,\ 6, \ 8$, one can find the complexities of the Iterative and FA approaches are, respectively, in the order of $\mathcal{O}\left(10^{4}\right)$ and $\mathcal{O}\left(10^{8}\right)$. When the number of antennas elements are large, letting $T=N=M_t$, it can be shown that the dominant terms of the complexities of the Iterative and FA approach are in the order of $\mathcal{O}\left(M_t^{4}\right)$ and $\mathcal{O}\left(M_t^{6}\right)$, respectively. The trade off between the power saving gain and computational complexity of the proposed FA approach in comparison with the Iterative method should be considered by the network designer/operator.
\begin{figure}[t]
\centering
    \includegraphics[width=.5\textwidth]{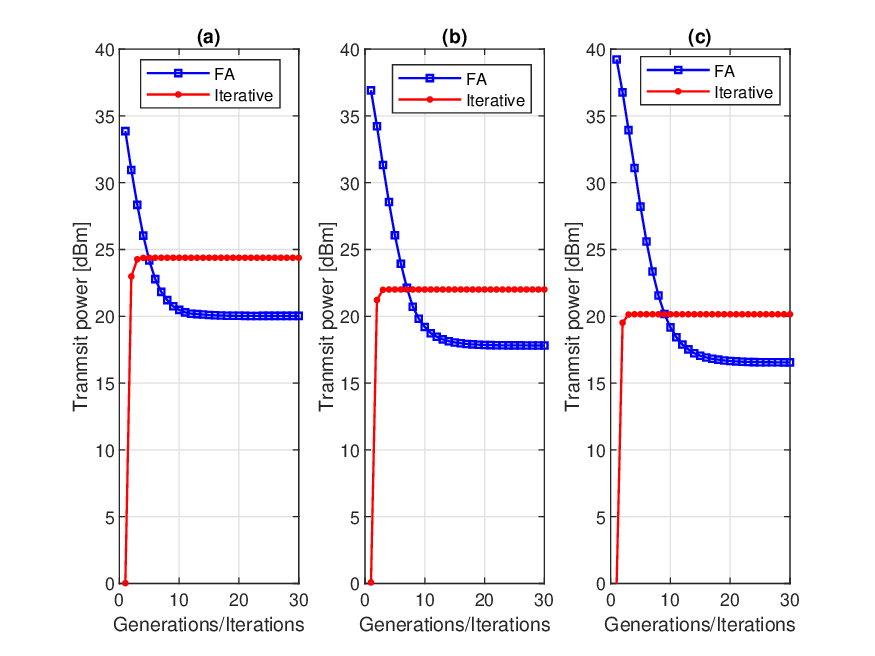}
\caption{ The total BS's transmit power versus the generations/iteration with different numbers of BS's antennas: (a) 4 antennas; (b) 6 antennas; (c) 8 antennas. The firefly population is $N=30$. The required SINR level at each user is $10$ dB.}\label{FATxBF_generation}
\end{figure}

Fig.~\ref{FATxBF_generation} shows the total BS's transmit power of the Iterative and proposed FA versus the number of iteration/generations with different numbers of BS's antennas. The results indicate that the Iterative approach converges after just 5 iterations/generations while the proposed FA requires about 20 generations/iterations to level off.

\begin{figure}[t]
\centering
   \includegraphics[width=.5\textwidth]{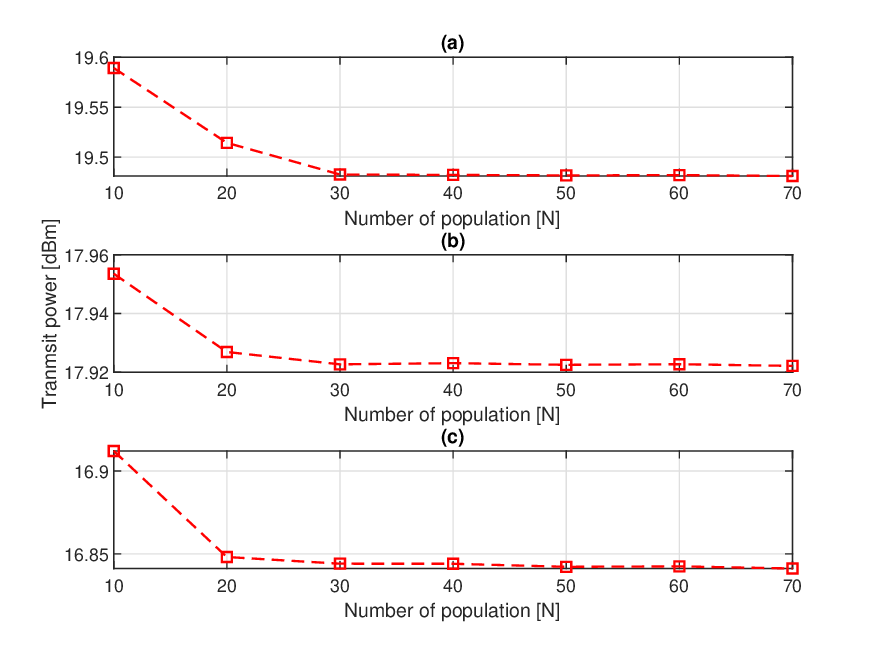}
\caption{ The total BS's transmit power versus the number of population with different numbers of BS's antennas: (a) 4 antennas; (b) 6 antennas; (c) 8 antennas. The number of generation is $T=30$. The required SINR level at each user is $10$ dB.}\label{FATxBF_population}
\end{figure}

Fig.~\ref{FATxBF_population} shows the total BS's transmit power of the proposed FA approach versus the number of population $N$ with different BS's antenna elements. It can be seen that the observed curves converge after $N=30$. Our simulations indicate that the proposed FA approach performs well with at least 30 fireflies to solve \eqref{TxBF_rearrange} under the investigated SINR range.
\subsection{Evaluations on Cognitive Transmit Beamforming}
We first reproduce the result of the experiment described in Example 1 of \cite{Yongwei} to compare the proposed FA approach with the SDP approach. In that experiment,
three SUs are located at $-5^{\circ}$,
$10^{\circ}$, $25^{\circ}$, and two PUs are located at
$30^{\circ}$ and $50^{\circ}$, relative to the BS's array broadside. The tolerable interference level two PUs are $I_{\textrm{to},1}=0.001$ and $I_{\textrm{to},2}=0.0001$. The noise variance is set to 0.1 while the required SINR values are set to 1 for the SUs. 

The channel covariance matrices from the secondary BS to SU $t$ , i.e.,
$\mathbf{R}_{s,t}=\mathbf{R}\left( \zeta_{s,t},\delta_a\right)$, and to PU $k$, i.e., $\mathbf{R}_{p,k}=\mathbf{R}\left( \zeta_{p,k},\delta_a\right)$, are the function of the angle of departure, i.e., $\zeta_{s,t}$ or $\zeta_{p,k}$, and the standard deviation of the angular spread, i.e., $\delta_a$. The $(m,n)$th entry of $\mathbf{R}\left(
\zeta,\delta_a\right)$ is, \cite{Mats}:
\begin{equation} \label{covaformula}
e^{\frac{j2\pi \Delta}{\psi}\left[\left(n-m\right)\text{sin}\zeta\right]} e^{-2\left[\frac{\pi \Delta \delta_a}{\psi}\left\{\left(n-m\right)\text{cos}\zeta\right\}\right]^2},
\end{equation}
where $\psi$ is the carrier wavelength, $\sigma_a=2^\circ$, and the antenna
spacing at the BS is set as $\Delta=\psi/2$. 

\begin{figure}[t]
\centering
    \includegraphics[width=.5\textwidth]{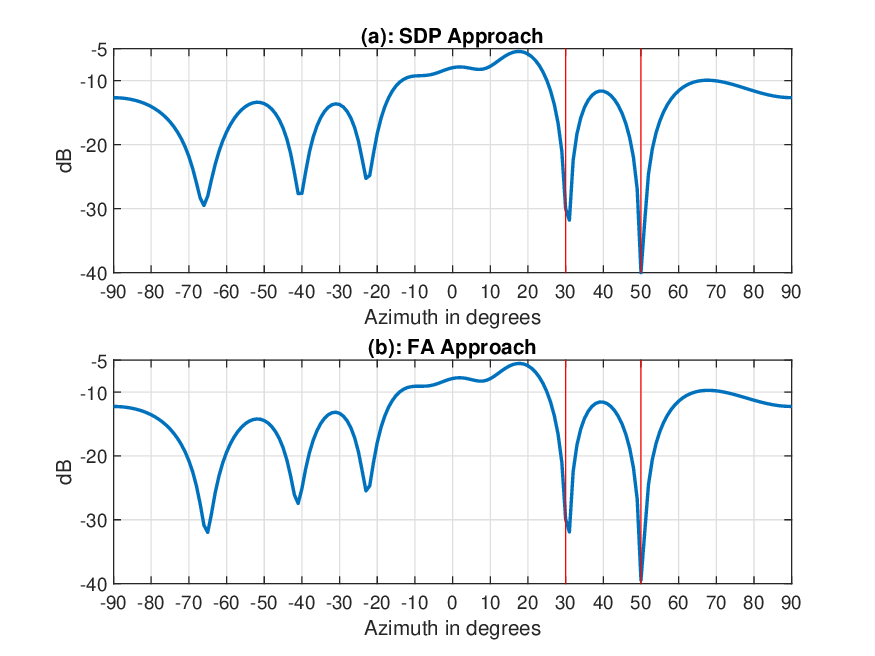}
\caption{ The radiation
pattern of the BS with 8 antennas: (a) The reproduction of \cite[Fig. 3]{Yongwei}; (b) The proposed FA approach with the number of population $N=100$.}\label{FAbeam}
\end{figure}
Fig.~\ref{FAbeam}~(a) illustrates the radiation patterns at the BS of the SDP approach as described in \eqref{cog_perfect_1}, which is the reproduction of Fig.~3 in \cite{Yongwei}, while Fig.~\ref{FAbeam}~(b) shows the radiation patterns at the BS of the FA approach proposed in Algorithm \ref{FAcognitive}. The results clearly indicate that the FA obtains the same radiation pattern as the SDP approach does. Both approaches are able to form nulls to the locations/angles where the PUs are located. In other words, the proposed FA can obtain the same optimal solution as the IPM does for the SDP counterpart. This confirms the ability of the proposed FA in handling highly nonlinear and multimodal optimization problems.

With the setup in Fig.~\ref{FAbeam}, i.e., $M_t=8$, $U=3$, $K=2$, $N=100$ and, $T=80$, one can easily verify from Lemmas~\ref{lemSDPcog} and \ref{lemFAcog} that the proposed FA approach requires higher computational complexity than the SDP approach does when it returns rank-one optimal solution. When the number of antennas is large, one can show that the dominant term of \eqref{Comp_SDPcog} is $M_t^{6\frac{1}{2}}$. On the other hand, assuming $T=N=M_t$, the dominant term of \eqref{Comp_FAcog} is $M_t^6$. Hence, the complexity of an IPM to solve \eqref{cog_perfect_1} is slightly higher than the complexity of the proposed FA in Algorithm~\ref{FAcognitive}, i.e., $\mathcal{O}\left( M_t^{6\frac{1}{2}}\right)$ in comparison with $\mathcal{O}\left(M_t^6\right)$.
\begin{figure}[t]
\centering
    \includegraphics[width=.5\textwidth]{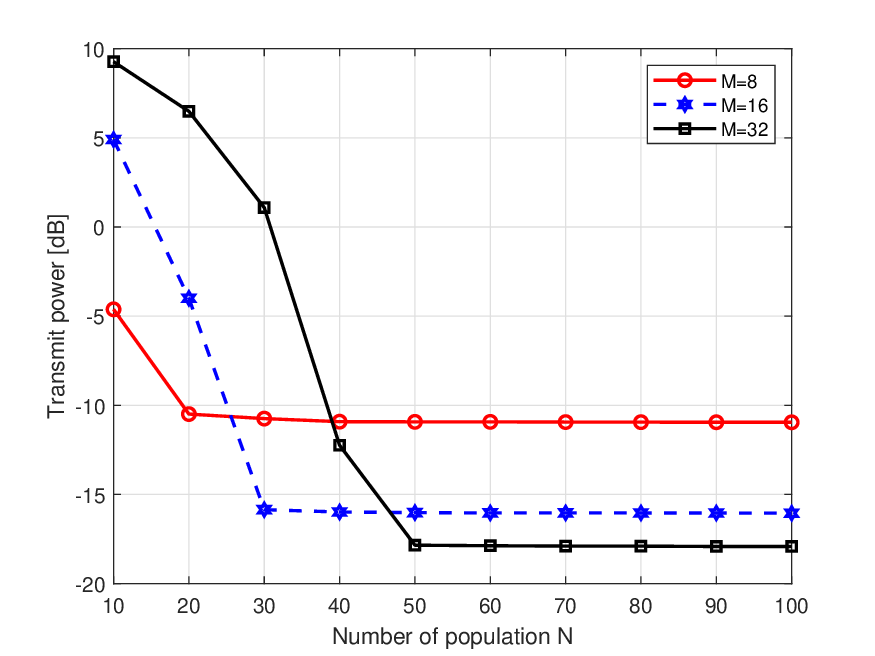}
\caption{ The total transmit power of the proposed FA approach versus the number of population with different numbers of transmit antennas.
The number of maximum generation $T=150$.} \label{Tx_N}
\end{figure}

Fig.~\ref{Tx_N} shows the transmit power of the proposed FA approach versus the number of population with different numbers of transmit antennas. The results indicate that the proposed FA converges with all number of antenna setups as all the observed curves level off after the maximum size of population of $N=50$. However, the higher of the antenna elements is, the larger the size of the population is required for a converged transmit power. For example, with $M=8$, 16, and 32, the proposed FA approach, respectively, obtains a stable transmit power at $N=30$, 40 and 50. This is due to the fact that the size of the system increases with a higher number of antenna elements, i.e., a higher degree of freedom. As a result, it requires a larger size of the population to provide a sufficient diversification for the exploration of the FA. The results also show that the required transmit power decreases when the number of antennas increase as the result of having higher degree of freedom.

\begin{figure}[t]
\centering
    \includegraphics[width=.5\textwidth]{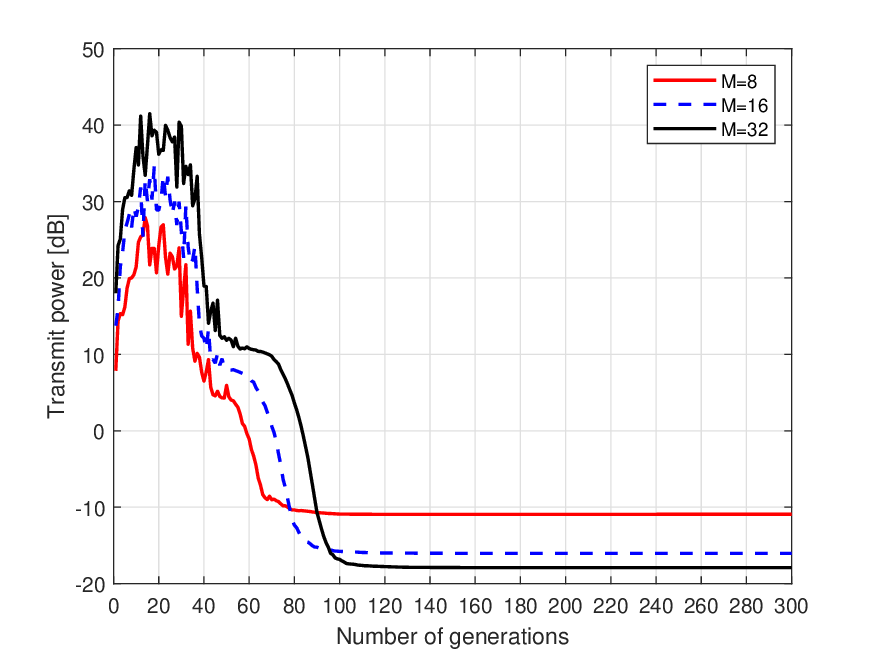}
\caption{ The total transmit power of the proposed FA approach versus the number of maximum generations with different numbers of transmit antennas. The number of population $N=70$.}\label{Tx_G}
\end{figure}

Fig.~\ref{Tx_G} depicts the transmit power of the proposed FA approach versus the number of maximum generations with different numbers of transmit antennas.  A similar trend as in Fig.~\ref{Tx_N} is also observed in this figure. The transmit power attained by the proposed FA approach converges with all numbers of antenna setups. The higher number of antennas is, the higher number of generations is needed as a result of higher exploitation required for the increase of the problem dimension. For instance, the transmit power levels off at around 90, 100, and 120 generations, respectively, for $M=8$, 16, and 32. 
\subsection{Evaluations on RIS-aided Transmit Beamforming}\label{ris_section}
We simulate a RIS-aided communication system which consists of one BS, one RIS, and two users, i.e., $U=2$. The distance between the BS and the RIS is $10$ m. Users are randomly distributed with a distance of $6$ m from the RIS. The pathloss exponents of both wireless links from the BS to the RIS and from the RIS to users are set to be 2.2 with the signal attenuation at the reference distance of 1 m being 30 dB \cite{Wu2020}, i.e., the large-scale fading coefficient is modeled as $-30 -22\log_{10}(d)$ dB where $d$ is the distance between the BS to RIS or RIS to a user. The noise variance at each user is $-124$~dBm. Monte Carlo simulations are carried over $100$ channel realizations. Each channel realization is associated with a random user location and a random fading coefficient.

\begin{figure}[t]
\centering
    \includegraphics[width=.5\textwidth]{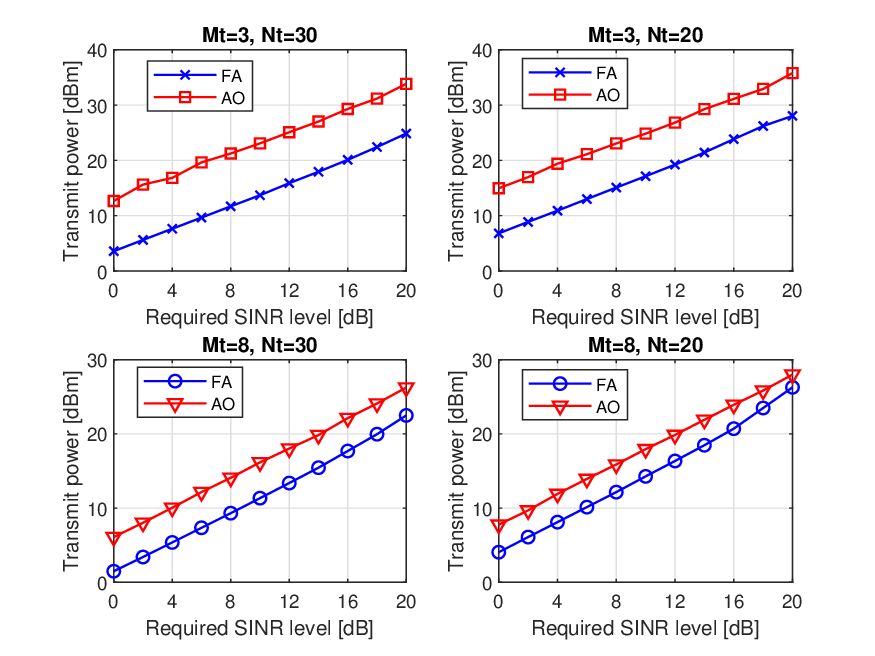}
\caption{ The total BS's transmit power versus the required SINR level with different numbers of BS's antennas and RIS's reflective elements. The firefly population is $N=120$. The number of maximum generations $T=50$.}\label{Tx_SINR}
\end{figure}

Fig.~\ref{Tx_SINR} illustrates the total BS's transmit power versus the required SINR level with different numbers of BS's antennas and RIS's reflective elements. The results indicate that the proposed FA prevails the AO approach in terms of lower power consumption. The superior performance of the FA approach over its AO counterpart can be explained as follows. As the AO approach approximates non-convex problem \eqref{IRS_PerfectCSI} by two convex sub-problems \eqref{prob_IRS_W_relax} and \eqref{prob_IRS_Theta_relax}, the solution obtained by the AO approach is not necessary the global optimal solution of the original problem \eqref{IRS_PerfectCSI}. On the other hand, the proposed FA possessing both exploitation and exploration abilities can effectively handle such non-convex problem and obtain much better solution than its counterpart. The results shown on Fig.~\ref{Tx_SINR} verify the ability of the proposed FA in handling highly nonlinear and multimodal optimization problems.

It can be observed from Fig.~\ref{Tx_SINR} that at a given number of RIS's reflective elements, the performance gap between the proposed FA and the AO decreases when the number of BS's antennas increases. For example, when $N_t=20$, the gaps are, respectively, around 7.5 dB and 3.5 dB with $M_t=3$ and $M_t=8$. Fortunately, at a given number of BS's antennas, the performance gap improves when the number of RIS's elements increases. For instance, with $M_t=8$, the performance gap increases from around 3.5 dB to 4.5 dB when $N_t$ increases from 20 to 30. Interestingly, the FA performs especially well with a relatively high ratio of $N_t/M_t$, i.e., the performance gap is around 9.5 dB with the ration of $30/3$ while it is around 3.5 with the ratio of $20/8$. The results can be explained as follows. A higher number of RIS's reflective elements gives more degree of freedom for the FA to perform. Moreover, the channel between the RIS and these users plays a higher role than that between the BS and the RIS does as the former is closer to these users. Last but not least, the performance gaps slightly decrease at relatively high SINR level especially when the $N_t/M_t$ ratio is relatively low. For example with the ratio of $20/8$, the performance gap is around 1.8 dB at SINR of 20 dB compared with around 3.5 dB at the other SINR levels, i.e., see the bottom-right corner figure of Fig.~\ref{Tx_SINR}. This is because of a fact that the FA has reached its limit of exploration with $N=120$ fireflies, at a stricter constraint condition.

We now compare the computational complexities of the AO and FA approaches for the experiments presented on Fig.~\ref{Tx_SINR}. As $N_t$ is larger than $M_t$, from Lemma~\ref{lemAO} one can show that the dominant term of the complexity of the AO approach is $n_0 N_t^{6\frac{1}{2}}$. Similarly, from Lemma~\ref{lem_RIS} one can conclude that the dominant term of the complexity of the FA approach is $TN^3N_t^2$. Substituting for $N_t=30$, $n_0=10$, $N=120$ and $T=50$, we can arrive at the fact that the computational complexities of the AO and FA approaches are on the same order of $\mathcal{O}\left( 10^{10}\right)$. When the numbers of antennas $M_t$ and $N_t$ are large, letting $N_t=n_0=M_t$ in \eqref{ComIMPRIS}, one can show that the dominant term of the complexity to attain $\varepsilon$-solution to \eqref{IRS_PerfectCSI} is $M_t^{7\frac{1}{2}}$.  On the other hand, one can derive the dominant term of \eqref{Comp_FARIS} as $M_t^6$ when assuming $T=N=N_t=M_t$. Hence, the complexity of an IPM to solve \eqref{IRS_PerfectCSI} is higher than the complexity of the proposed FA in Algorithm~\ref{FARIS}, i.e., $\mathcal{O}\left( M_t^{7\frac{1}{2}}\right)$ in comparison with $\mathcal{O}\left(M_t^6\right)$. 
\begin{figure}[t]
\centering
    \includegraphics[width=.5\textwidth]{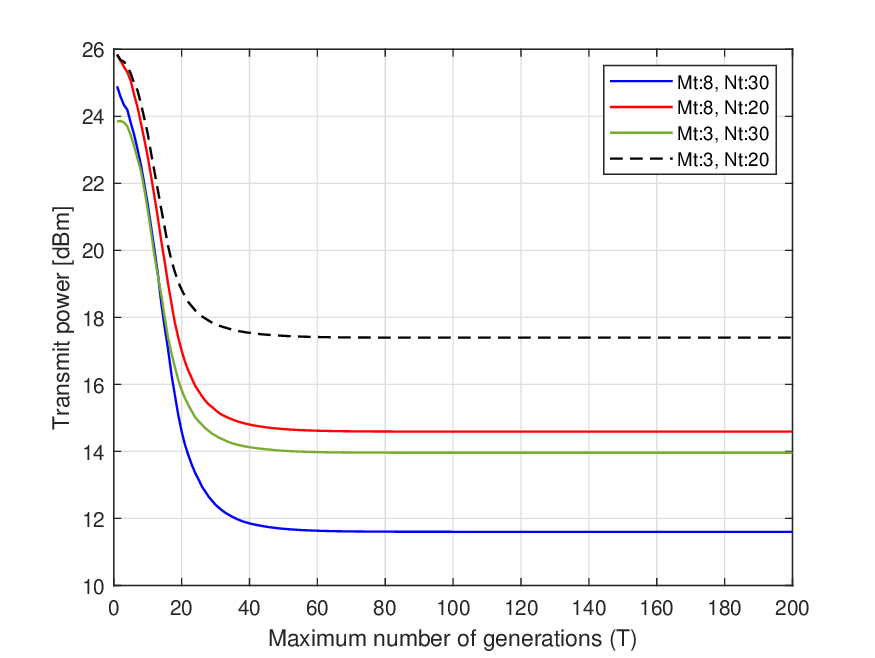}
\caption{ The total BS's transmit power versus the number of maximum generations with different numbers of BS's antennas and RIS's reflective elements. The firefly population is $N=120$. The required SINR level is 10 dB.}\label{RIS_Tx_generation}
\end{figure}

In Fig.~\ref{RIS_Tx_generation}, the total BS's transmit power is plotted versus the maximum of generation $T$ used in the FA in Algorithm~\ref{FARIS} with different BS's antennas and RIS's reflective elements. The results indicate that the proposed FA requires around $50$ to $60$ generations to attain the optimal solution for all setups.

\begin{figure}[t]
\centering
    \includegraphics[width=.5\textwidth]{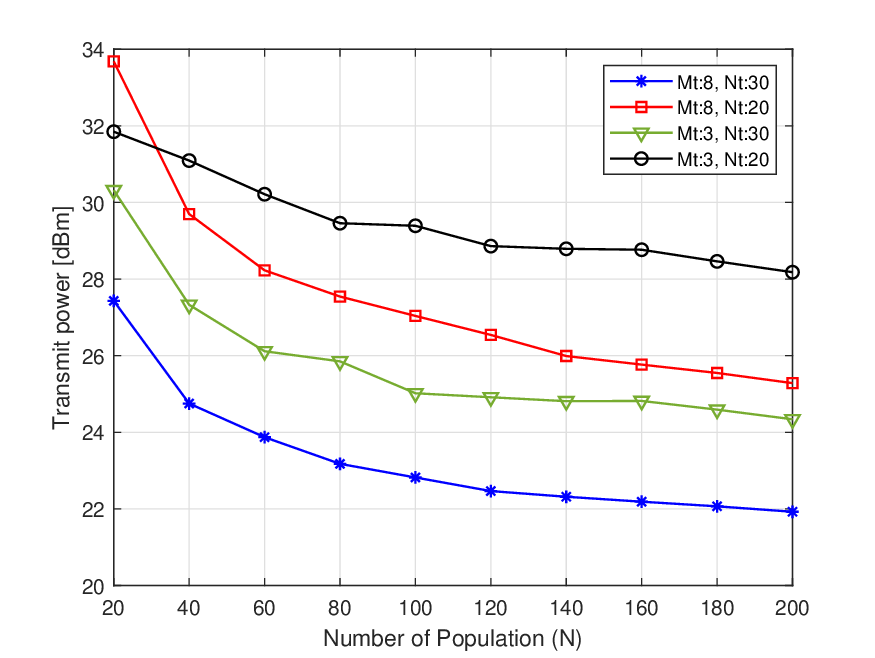}
\caption{ The total transmit power versus the number of populations with different numbers of BS's antennas and RIS's reflective elements. The number of maximum generations $T=50$. The required SINR level is 20 dB.}\label{RIS_Tx_population}
\end{figure}

Fig.~\ref{RIS_Tx_population} illustrates the total transmit power versus the number of population $N$ with different BS's antennas and RIS's elements. The results show that increasing the size of the firefly population enables the FA to obtain better solution. For example, the total transmit power decreases around $7$ dB, $5.4$ dB, $5$ dB, and $3$ dB, respectively, for the setups of $(M_t=8, N_t=20)$, $(M_t=3, N_t=30)$, $(M_t=8, N_t=20)$, and $(M_t=3, N_t=20)$ when the firefly population increases from $20$ to $120$. The performance gap at the 20 dB SINR level observed in Fig.~\ref{Tx_SINR} for  $(M_t=8, N_t=20)$ can be improved 1 dB further  when the population size is enlarged from 120 to 200. These total-transmit-power curves converge after $N=180$ as the reduction in the total transmit power is negligible  when the population increases to $N=200$ for all setups.
\subsection{Evaluations on RIS-aided WPT}
Here, we use the same setup for the RIS-aided communication system as considered in the previous section, i.e., Section \ref{ris_section}. However, the EHRs are randomly placed with the distance of 2 m from the RIS. We run $m_0=10$ iterations to obtain the solution for the SCA approach. 
\begin{figure}[t]
\centering
    \includegraphics[width=.5\textwidth]{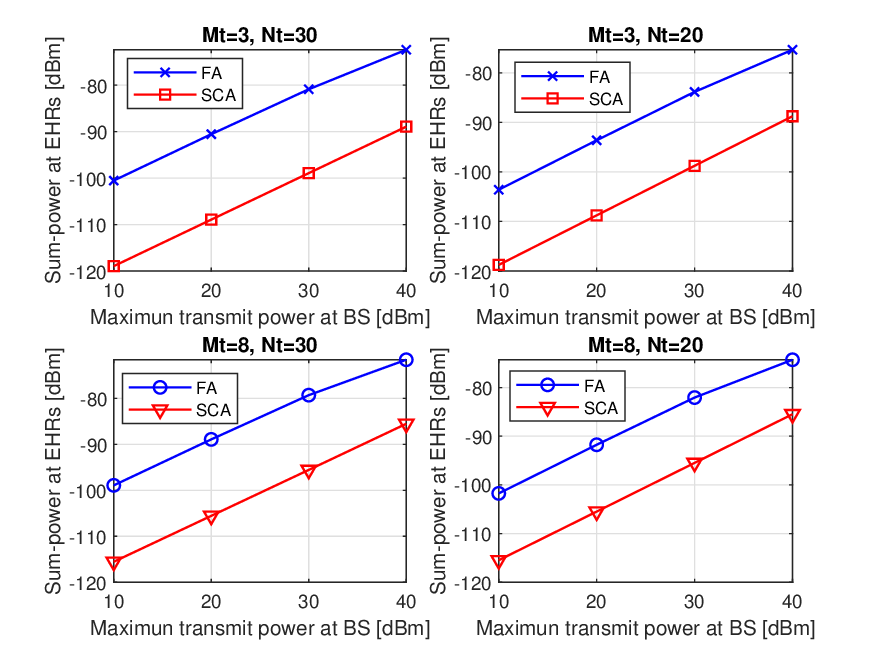}
\caption{ Sum-power received at EHRs versus BS's maximum transmit power with different numbers of BS's antennas and RIS's reflective elements. The firefly population is $N=100$. The number of maximum generations $T=50$.}\label{Ex_Po}
\end{figure}

Fig.~\ref{Ex_Po} shows the sum-power received at EHRs versus BS's maximum transmit power with different numbers of BS's antennas and RIS's reflective elements. It is clear from the figure that the proposed FA approach outperforms the SCA approach in \cite{Wu2020} in offering higher sum-power at EHRs. The performance gaps are, respectively, around $18$ dB, $17$ dB, $15$ dB, and $14$ dB for the setups of $(M_t=3, N_t=30)$, $(M_t=8, N_t=30)$, $(M_t=3, N_t=20)$, and $(M_t=8, N_t=20)$. The superior performance of the proposed FA over the SCA is due to the advantage of having exploitation and exploration abilities to handle non-convex optimization problems. On the other hand, the SCA employs the first-oder Taylor expansion to approximate the optimization problem resulting in a lower-bounded solution. Furthermore, the FA approach allocates one active beamforming vector for each EHR whereas the SCA only uses one active beamforming vector for all EHRs.  The results shown on Fig.~\ref{Ex_Po} again verify the ability of the proposed FA in handling highly nonlinear and multimodal optimization problems.

Comparing Figs.~\ref{Tx_SINR} and \ref{Ex_Po}, it can be observed that the FA behaves in a similar manner for both power minimization problem \eqref{IRS_PerfectCSI_n} and sum-power maximization problem \eqref{IRS_WPT_n}. For instance, at the same value of $M_t$, the higher the value of $N_t$, the larger the performance gap is. At the same value of $N_t$, the lower the value of $M_t$, the bigger the performance gap is. The results also recommend to maintain a relatively high ratio of $N_t/M_t$ to attain the best performance of the FA. Slight declines in the performance gaps are also observed at the stricter constraint of BS's transmit power, i.e., 40 dBm, as the FA's population reach their limit of exploration.
\begin{figure}[t]
\centering
    \includegraphics[width=.5\textwidth]{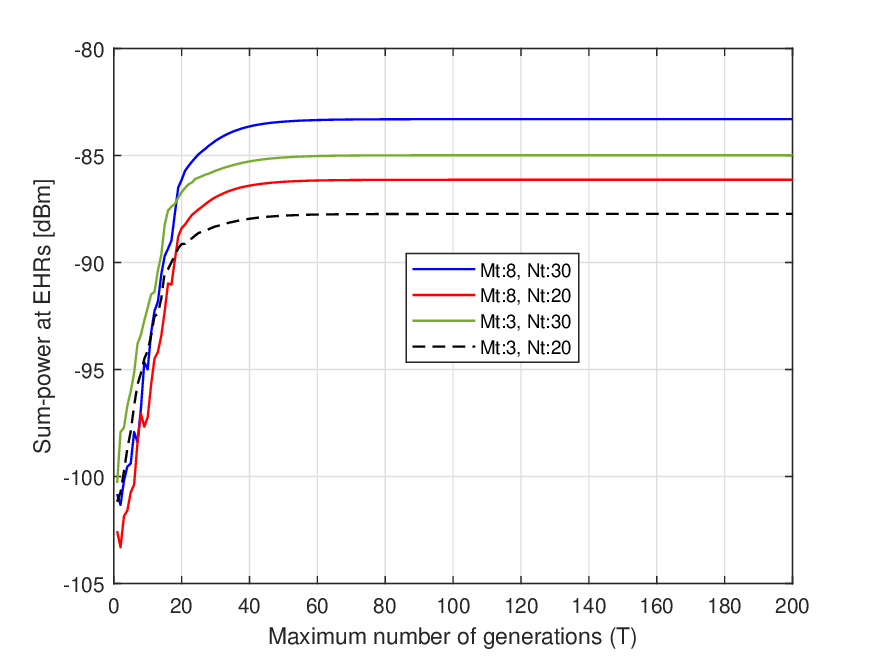}
\caption{ Sum-power received at EHRs versus the number of maximum generations with different numbers of BS's antennas and RIS's reflective elements. The firefly population is $N=100$. The required SINR level is 10 dB.}\label{RIS_Ex_generation}
\end{figure}

We proceed by comparing the computational complexities of the SCA and FA approaches for the experiments shown on Fig.~\ref{Ex_Po}. As $N_t$ is larger than $M_t$, from Lemmas~\ref{lem_SCA} and \ref{lem_FARIS_WPT}, it is clear that the dominant terms of the complexities of the SCA and the FA approaches are, respectively, $m_0 N_t^{3}$ and $TN^3N_t^2$. Substituting for $N_t=30$, $m_0=10$, $N=100$ and $T=50$, we can arrive at the fact that the computational complexities of the SCA and FA approaches are, respectively, on the orders of $\mathcal{O}\left( 10^{5}\right)$ and $\mathcal{O}\left( 10^{10}\right)$. When the numbers of antennas $M_t$ and $N_t$ are large, letting $N_t=m_0=M_t$ in \eqref{ComSCA}, one can show that the dominant term of the complexity of the SCA is $M_t^{4}$.  On the other hand, the dominant term of \eqref{Comp_WPT} is $M_t^6$ when assuming $T=N=N_t=M_t$. Hence, the complexity of the SCA approach is lower than that of the proposed FA in Algorithm~\ref{FARIS_WPT}, i.e., $\mathcal{O}\left( M_t^{4}\right)$ in comparison with $\mathcal{O}\left(M_t^6\right)$. 

Sum-power received at EHRs are shown versus the number of maximum generations with different numbers of BS's antennas and RIS's reflective elements in Fig.~\ref{RIS_Ex_generation}. The figure reveals that the proposed FA converges after around $50$ to $60$ generations for all observed setups.
\begin{figure}[t]
\centering
    \includegraphics[width=.5\textwidth]{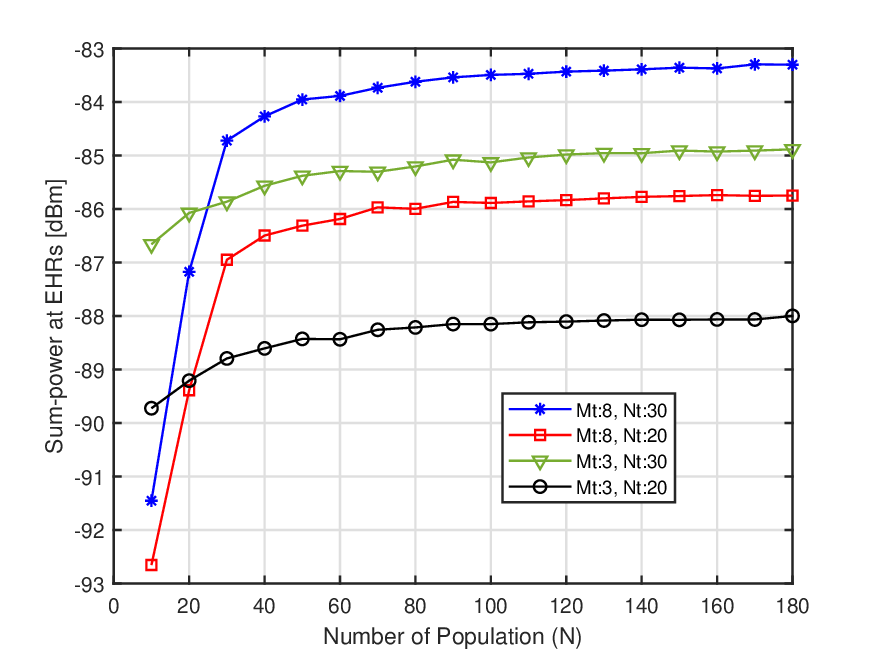}
\caption{ Sum-power received at EHRs versus the number of populations with different numbers of BS's antennas and RIS's reflective elements. The number of maximum generations $T=50$. The required SINR level is 20 dB.}\label{RIS_Ex_population}
\end{figure}

The effect of the firefly population on the sum-power received at EHRs is illustrated on Fig.~\ref{RIS_Ex_population}. The figure shows that all the curves converge after the population size of 80. However the difference between the EHRs' sum-power offered by 80 fireflies and that offered by 40 fireflies is no more than 0.7 dB for all observed setups. This indicates that the complexity of the proposed FA for the RIS-aided WPT sum-power maximization problem in  \eqref{IRS_WPT_n} can be reduced with an acceptable tradeoff in the optimality.
\section{Conclusion}
We have proposed a generalized FA to find optimal solution for an optimization framework containing objective function and constraints as multivariate functions of independent optimization variables. We have adopted the proposed generalized FA to solve four representative examples of classic transmit beamforming, cognitive beamforming, RIS-aided transmit beamforming, and RIS-aided wireless power transfer. Our analyzes have indicated that the computational complexities of proposed FA approaches are less than those of their IPM counterparts, i.e., the SDP and the AO approaches, yet higher than that of the iterative and SCA approaches in large-antenna scenarios. Simulation results have revealed the fact that the proposed FA attains the same optimal solution as the IMP does for the under-investigated cognitive beamforming problem. Interestingly, the proposed FA outperforms the iterative, AO, and SCA approaches for the under-investigated classic transmit beamforming, RIS-aided transmit beamforming, and wireless power transfer problems, respectively. This confirms the effectiveness of the proposed generalized FA in handling multivariate and non-convex problems.
\bibliographystyle{IEEEtran}
\linespread{1.6}
\bibliography{TW_Nov_22_1723_Camera_Ready}
\begin{IEEEbiography}[{\includegraphics[width=1in,height=1.25in]{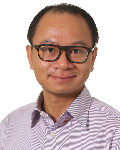}}]{Tuan Anh Le (S'10-M'13-SM'19)} received the Ph.D. degree in telecommunications research from
King’s College London, The University of London, U.K., in 2012. He was a Post-Doctoral Research Fellow with the School of Electronic and
Electrical Engineering, University of Leeds, Leeds, U.K. He is a Senior Lecturer at Middlesex University, London, U.K. His current research interests include integrated sensing and communication (ISAC), RIS-aided communication, RF energy harvesting and wireless power transfer, physical-layer security, nature-inspired optimization, and applied machine learning for wireless communications. He severed as a Technical Program Chair for 26th International Conference on Telecommunications (ICT 2019). He was an Exemplary Reviewer of IEEE Communications Letters in 2019.
\end{IEEEbiography}
\begin{IEEEbiography}[{\includegraphics[width=1in,height=1.25in]{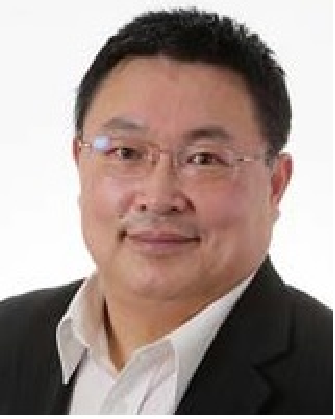}}]{Xin-She Yang} obtained his DPhil in Applied Mathematics from the
University of Oxford. He then worked at Cambridge University and
National Physical Laboratory (UK) as a Senior Research Scientist.
Now he is Reader at Middlesex University London, and a co-Editor
of the Springer Tracts in Nature-Inspired Computing. He is also
an elected Fellow of the Institute of Mathematics and its
Applications. He was the IEEE Computational Intelligence
Society (CIS) chair for the Task Force on Business Intelligence
and Knowledge Management (2015 to 2020). He has published
more than 300 peer-reviewed research papers with more
than 84,000 citations, and he has been on the prestigious
list of highly-cited researchers (Web of Sciences) for
eight consecutive years (2016-2023).
\end{IEEEbiography}
\end{document}